\documentclass[10pt]{article}

\usepackage{graphicx}
\usepackage{url}
\usepackage{times}
\usepackage{lscape}
\usepackage{cite}

\begin{document}
\pagestyle{empty}
\title{Experimental study of pedestrian counterflow in a corridor}

\author{Tobias Kretz, Anna Gr\"unebohm, Maike Kaufman,\\ Florian Mazur and Michael Schreckenberg\\ \\
Physik von Transport und Verkehr, Universit\"{a}t Duisburg-Essen,\\  47048 Duisburg, Germany\\ \\
\{kretz,gruenebohm,kaufman,mazur,schreckenberg\}\\@traffic.uni-duisburg.de\\
}

\maketitle

\begin{abstract}
In this work the results of a pedestrian counterflow experiment in a corridor of a width of $2$ meter are presented. 
$67$ participants were divided into two groups with varying relative and absolute size and walked in opposite direction through a corridor. 
The video footage taken from the experiment was evaluated for passing times, walking speeds, fluxes and lane-formation including symmetry breaking. 
The results include comparatively large fluxes and speeds as well as a maximal asymmetry between left- and right-hand traffic.
The sum of flow and counterflow in any case turns out to be larger than the flow in all situations without counterflow.
\end{abstract}

\section{Introduction}

Pedestrian counterflow may occur in a number of situations: in rather narrow corridors on board of ships; in shopping areas during Christmas time; or at pedestrian traffic lights and most importantly in emergency situations if there are no separate routes for rescue forces and evacuees.
Such situations may vary in the relative group size and differ in the time the counterflow situation exists: near equilibrium over comparatively large times in shopping areas while only for a few seconds at traffic lights.
One-directional pedestrian flow has often been investigated and summarized into one common fundamental diagram \cite{Weidmann92}.
For bi-directional pedestrian flow (counterflow) much less data are available \cite{Isobe04,Nagai05b}.
Yet there is a number of theoretical analyses \cite{Helbing95,Fukui99,Muramatsu99,Helbing01,Burstedde01,Burstedde01b,BlueAdler,PED01:Blue,Algadhi01,Tajima02,Kretz06}, and the study of counterflow is not limited to pedestrian counterflow, for example counterflow has recently attracted some interest on ant trails \cite{John04}, the dynamics of motor molecules \cite{Klumpp04}, and basic theoretical investigations \cite{Evans95}.

\section{The Scenario}
\begin{landscape}
\begin{figure}[p]
\begin{center}
\includegraphics[width=550pt]{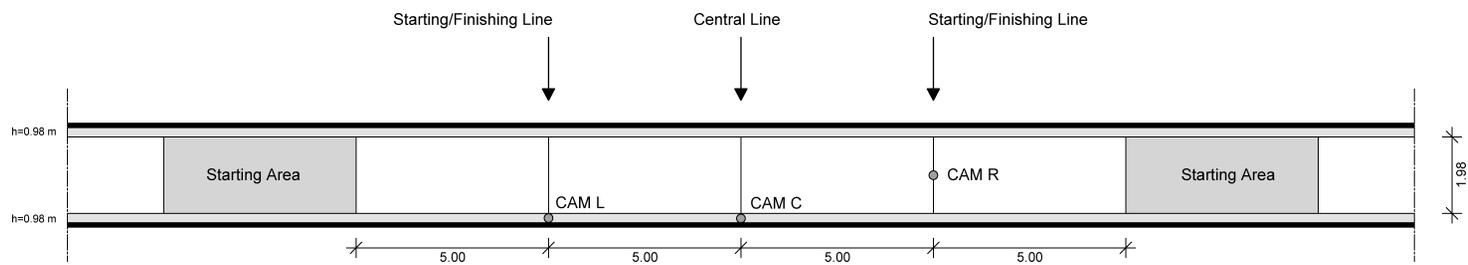}
\caption{The floor plan. The participants initially stood within the ``Starting Area", without further advises how they should arrange there. The positions of the cameras are marked with ``CAM L", ``CAM C", and ``CAM R", of which the latter one filmed from above and the other ones from the side. The distance between the cameras is $5$ meter., the height of the cornice $0.98$ meter and the width (without the extra space above the cornice) is $1.98$ meter.}
\label{fig:floorplan}
\end{center}
\end{figure}
\end{landscape}

\begin{figure}[htbp]
\begin{center}
\includegraphics[width=250pt]{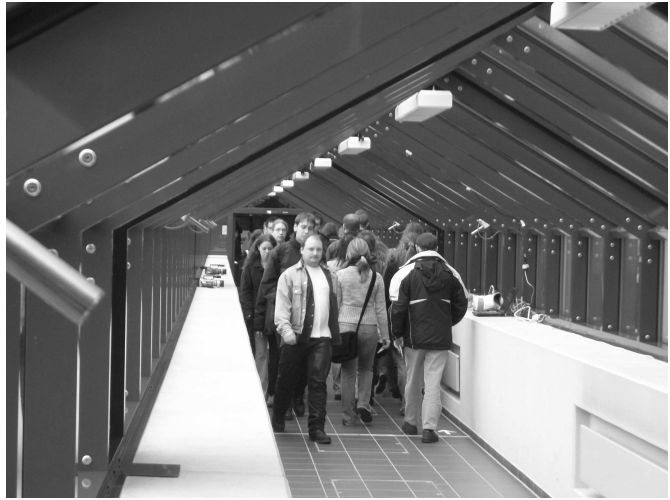}
\caption{A snapshot from the experiment.}
\label{fig:snapshot}
\end{center}
\end{figure}

The experiment, which follows test scenario $8$ of \cite{MSC1033}, took place at the Sportschule Wedau in Duisburg and is one out of two experiments (compare \cite{Kretz06b}) which have been conducted following \cite{MSC1033}.
The floor plan (see figure \ref{fig:floorplan} and \ref{fig:snapshot}) consisted of a corridor of a width of 1.98 meters and a length of about $34$ meters. 
$98$ cm above the ground the corridor broadened by $40$ cm on each side, increasing the effective width slightly as the participants could lean above this cornice to some extent. 
Three cameras with frame rates of $25$ respectively $30$ frames per second were used to obtain data.
The one to the left and to the right were placed at a distance of $5$ m next to the central camera. 
Two groups started walking five more meters outside the camera region after some acoustic signal.

\subsection{The Participants}
The majority of the $67$ participants ($33$ male and $34$ female) were students of Duisburg-Essen University, mostly born in the eighties.

\subsection{Group Sizes}
The amount of counterflow, which is always given as a size of the counter-group of the group in focus by the total number of participants, always was (approximately) $0$, $0.1$, $0.34$, or $0.5$. 
Those values were not always met exactly due to rounding errors and participants needing to pause. \par
While there were repeated runs with identical or very similar group size combinations, the assignment of the people to the two groups was changed each time.

\section{Results}
The resulting data (passing time, walking speed and specific flux) will now be given, either in dependence of the group size for (approximately) identical counterflow fractions and in dependence of (exact) counterflow fractions.
\subsection{Passing Times}
\begin{figure}[htbp]
\begin{center}
\includegraphics[width=250pt]{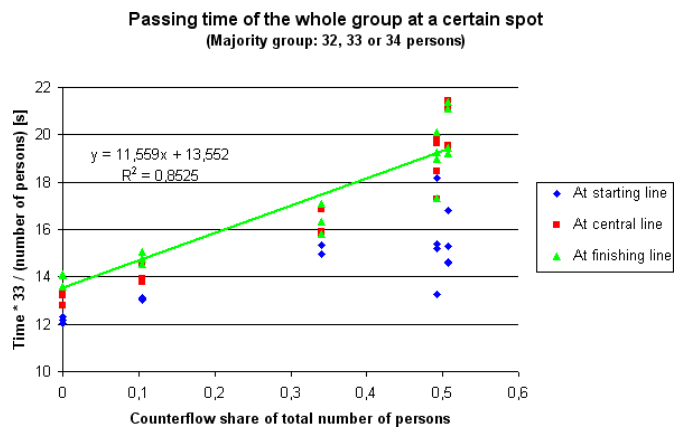}\\
\caption{Comparison of passing times of a majority group with approximately constant group size. Since the size of the majority group varied slightly, the times were scaled accordingly. This figure demonstrates the significant influence some counterflow has on the passing time. However, the influence is not that big as one might assume at first: imagining the group to occupy a rectangle with the width of the corridor in the case of no counterflow and half of the width of the corridor in the case of $0.5$ counterflow, one might guess, that the passing time at a spot behind the central meeting point of the two groups doubles. In fact it only increases by roughly $43$\%.}
\label{fig:comp_passing_times}
\end{center}
\end{figure}

The passing time is defined as the time a group needs to pass a certain spot.
\begin{figure}[htbp]
\begin{center}
\includegraphics[width=250pt]{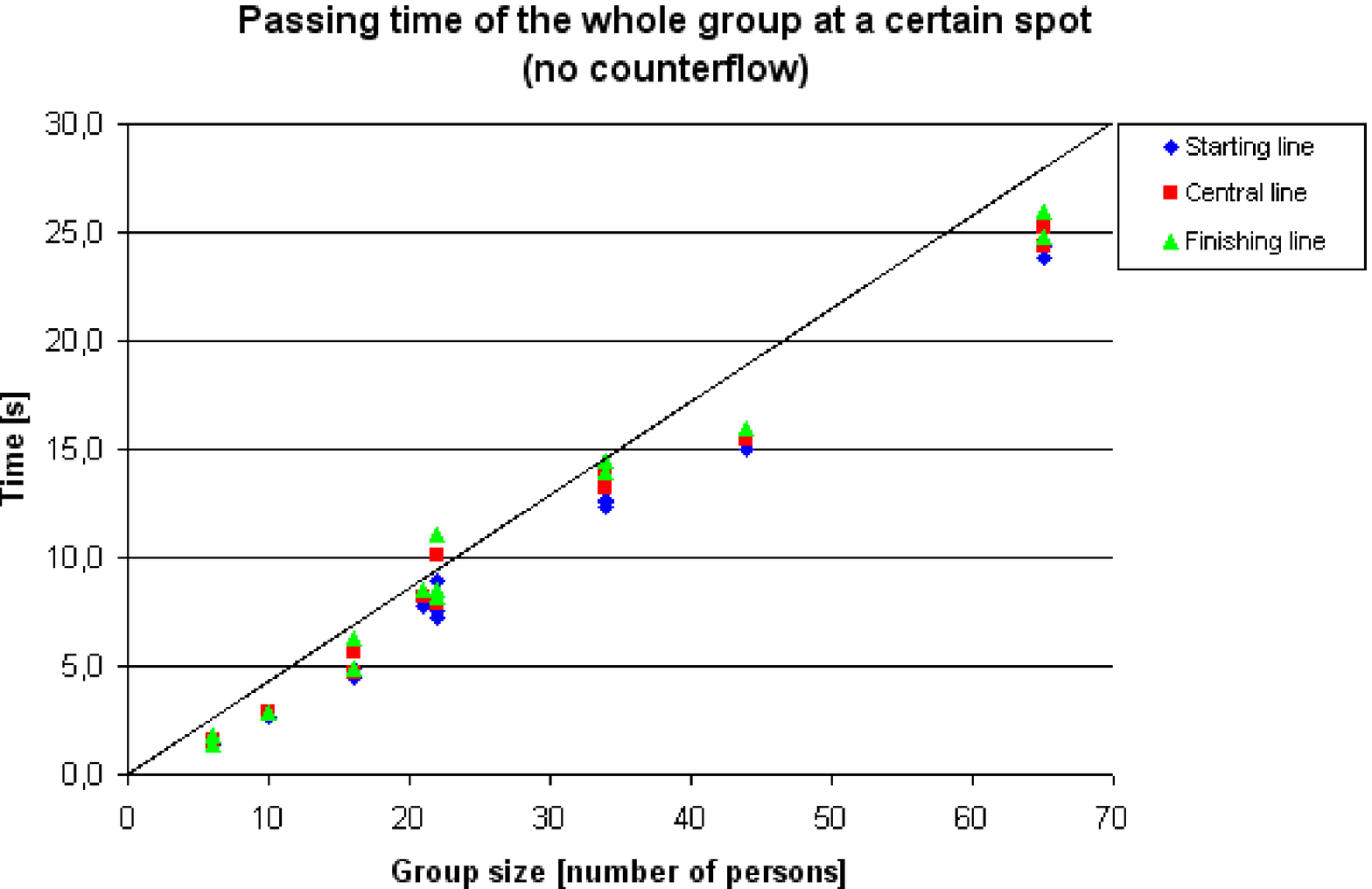}\\ \vspace{12pt}
\includegraphics[width=250pt]{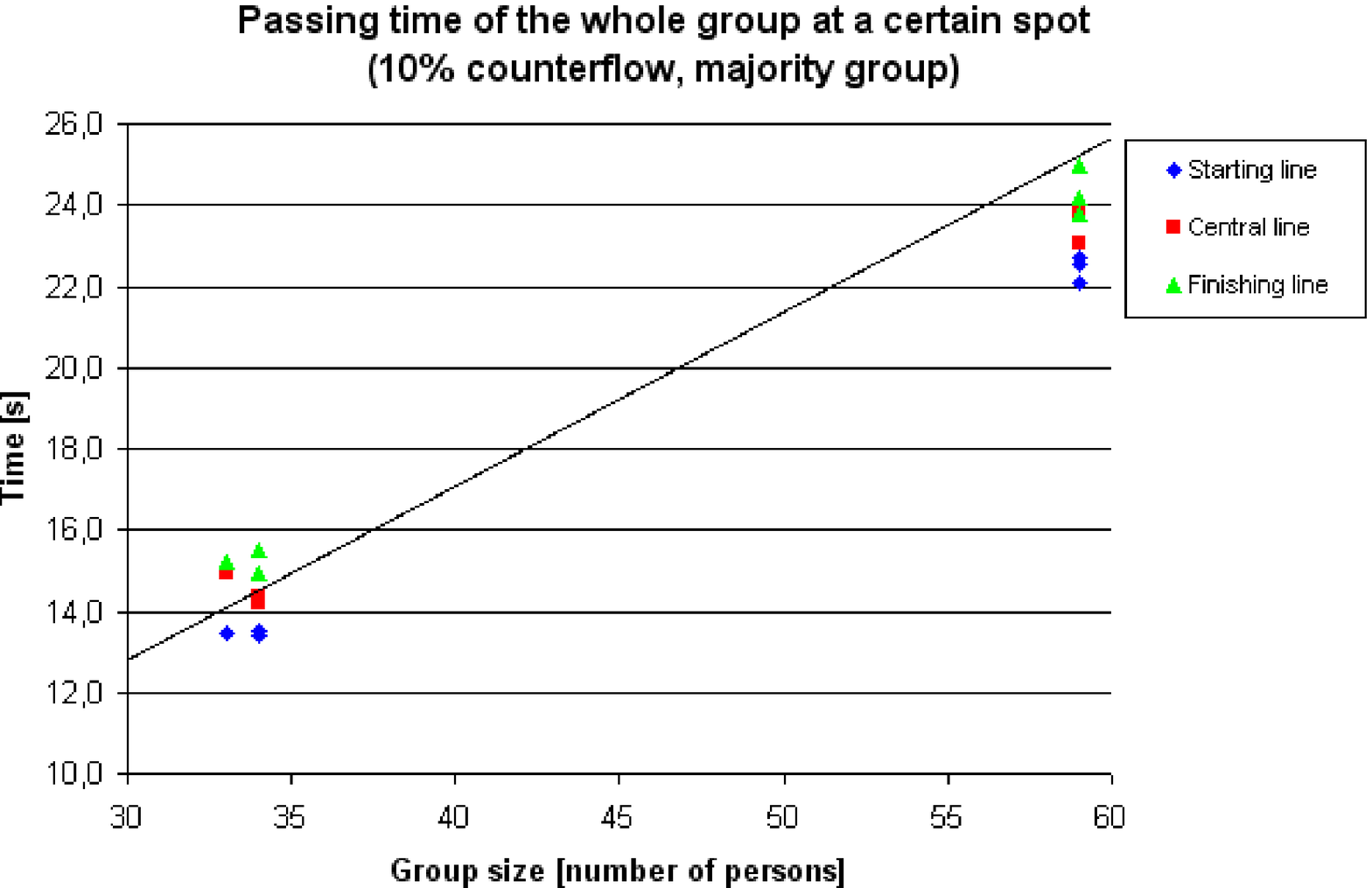}  \\ \vspace{12pt}
\includegraphics[width=250pt]{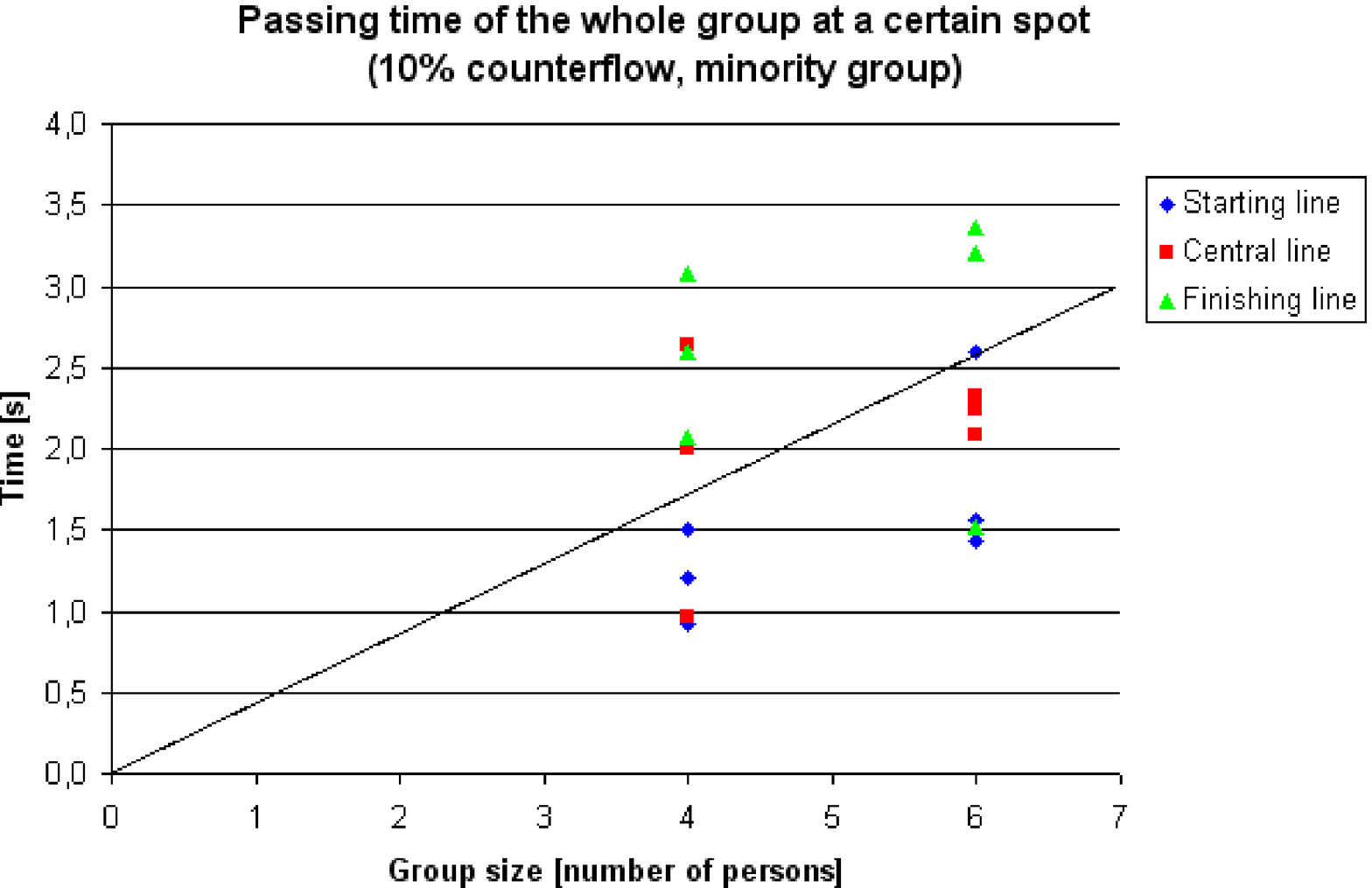} \\
\caption{Passing times (part $i$). The first thing one observes in this and figure \ref{fig:passing_timesB} is that the results vary significantly more in the presence of counterflow. Naturally this effect is most distinctive when the group is very small and the selection of the group members (e.g. due to varying body height) becomes important. It is a priori not obvious that for a counterflow situation the passing time increases linearly with the group size, as is in the case of no counterflow. Yet the results of the $50$\% counterflow situations (compare figure \ref{fig:passing_timesB}) justify this assumption and thus table \ref{tab:grad}.}
\label{fig:passing_timesA}
\end{center}
\end{figure}
\begin{figure}[htbp]
\begin{center}
\includegraphics[width=250pt]{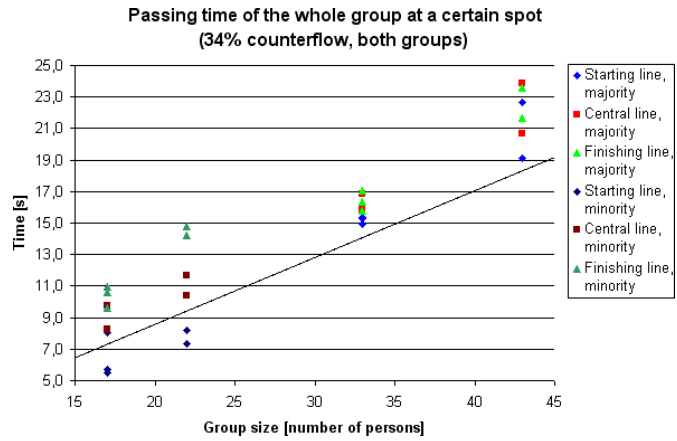}  \\ \vspace{12pt}
\includegraphics[width=250pt]{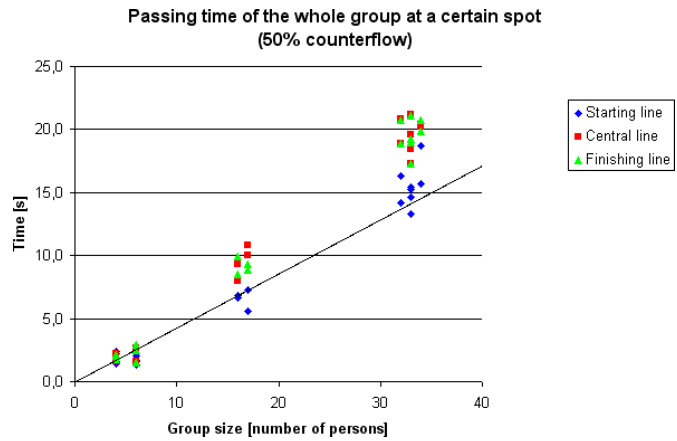} \\
\caption{Passing times (part $ii$).}
\label{fig:passing_timesB}
\end{center}
\end{figure}
As would have been expected, for each counterflow fraction the passing time grows linearly with the group size. See figures \ref{fig:passing_timesA} and \ref{fig:passing_timesB}.
More interesting is a comparison of the gradients of the different counterflow fractions at the finishing line (see table \ref{tab:grad}), which have the dimension of an inverse flux.
The large dispersion of the results for the minority group in the case of a counterflow fraction of $0.1$ follows from the small group size and from the important role which the choice of individuals plays that form the minority group.
 \begin{table}[htbp]
\begin{center}
\begin{tabular}{c|c|c|c|c}
&\multicolumn{2}{c|}{Offspring included}&\multicolumn{2}{c}{Offspring not included}\\
Fraction of counterflow & Dependence & $R^2$ & Dependence & $R^2$\\ \hline
0.00 & 0.39 $s$/pers.$\cdot n$ & 0.98 & 0.40 $s$/pers.$\cdot n$ $-$ 0.28 $s$ & 0.98\\ 
0.10 & 0.42 $s$/pers.$\cdot n$ & 0.96 & 0.36 $s$/pers.$\cdot n$ $+$ 3.20 $s$ & 0.99\\ 
0.34 & 0.51 $s$/pers.$\cdot n$ & 0.92 & 0.62 $s$/pers.$\cdot n$ $-$ 4.14 $s$ & 0.95\\ 
0.50 & 0.59 $s$/pers.$\cdot n$ & 0.98 & 0.62 $s$/pers.$\cdot n$ $-$ 0.99 $s$ & 0.99\\ 
0.66 & 0.64 $s$/pers.$\cdot n$ & 0.90 & 0.81 $s$/pers.$\cdot n$ $-$ 3.43 $s$ & 0.95\\ 
0.90 & 0.51 $s$/pers.$\cdot n$ &-0.49 & 0.06 $s$/pers.$\cdot n$ $+$ 2.36 $s$ & 0.01\\ 
\end{tabular}
\caption{Results of linear regressions for the dependence between passing time at the finishing line and group size, with and without forced inclusion of the offspring. $n$: number of persons. Note that a counterflow fraction of $0.90$ denotes the minority group of an experiment with $10$\% counterflow. Compare figures \ref{fig:passing_timesA} and \ref{fig:passing_timesB}.}
\label{tab:grad}
\end{center}
\end{table}
Another interesting result is the comparison of passing times for constant majority group size in figure \ref{fig:comp_passing_times}. The increase of the passing time even at the starting line from $0$ to $0.1$ counterflow shows that the participants reacted quite early to even just a few people approaching.

\begin{figure}[htbp]
\begin{center}
\includegraphics[width=250pt]{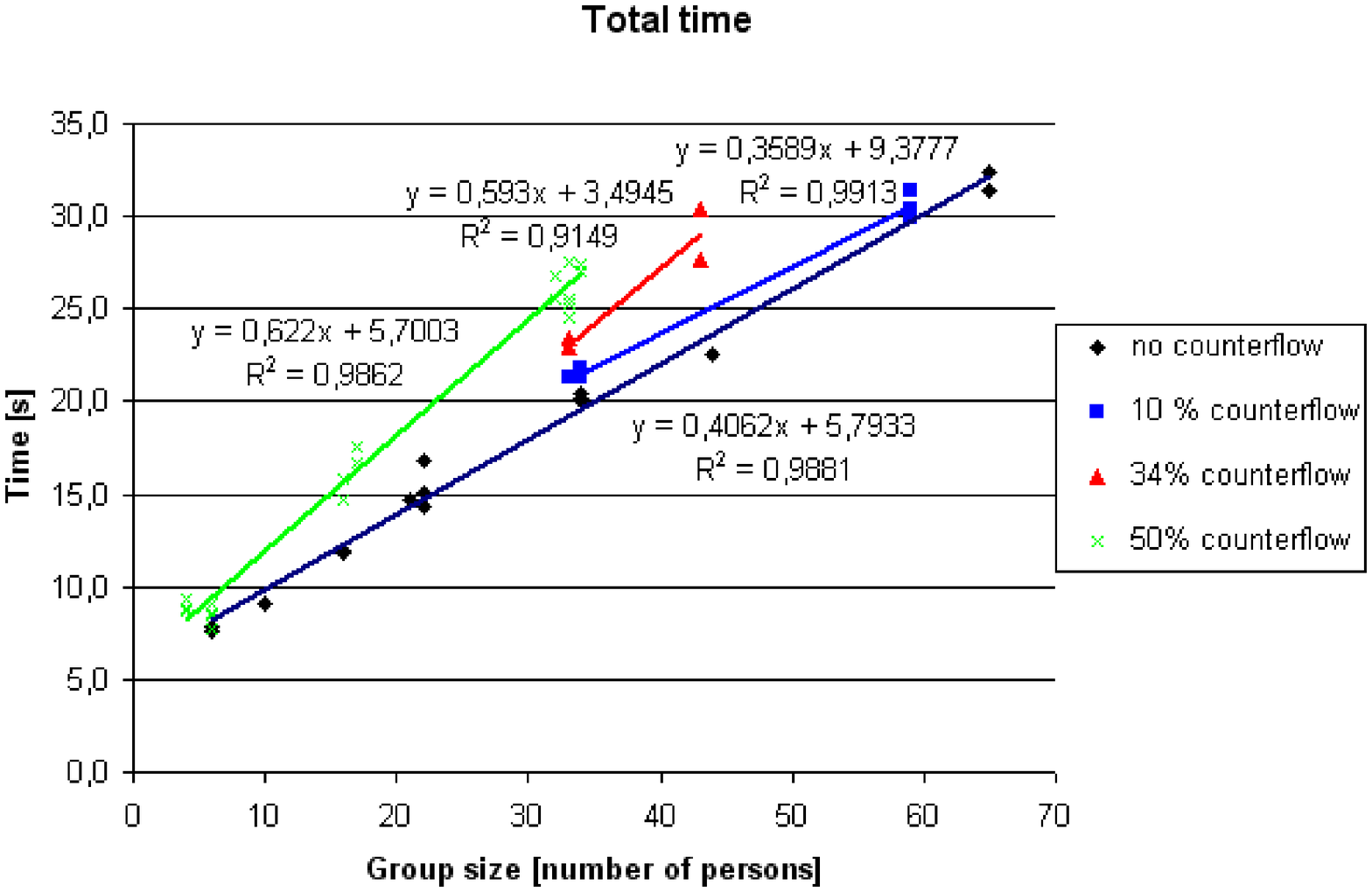}  \\ \vspace{12pt}
\includegraphics[width=250pt]{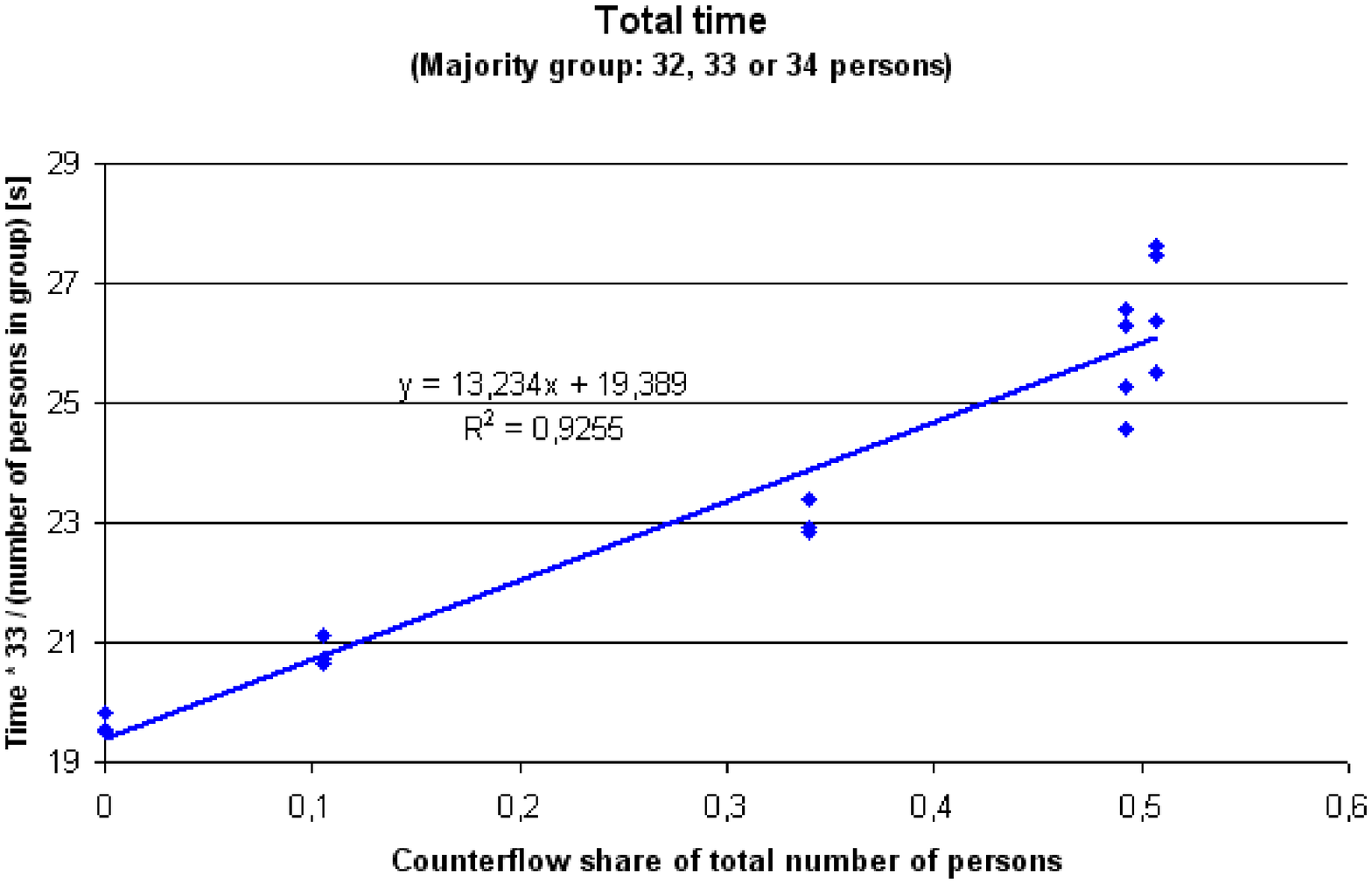}\\
\caption{Comparison of total times: The total time is the time between the passing of the starting line by the first person until the time of the passing of the finishing line by the last person. The first figure shows the total times of the majority group. The second figure exhibits an influence of the counterflow on the total time, which with an increase of approximately $34$\% from no to $0.5$ counterflow is slightly smaller than the influence on the passing time (compare figure \ref{fig:comp_passing_times}). The absolute value of the slope of the regressionline, however, is bigger than in the case of the passing times. The reason for this is, that there is a minimum time larger than zero for the total time, while the passing time can get very close to zero for very small groups.}
\label{fig:comp_total_times}
\end{center}
\end{figure}

\subsection{Walking Speeds}
\begin{figure}[htbp]
\begin{center}
\includegraphics[width=250pt]{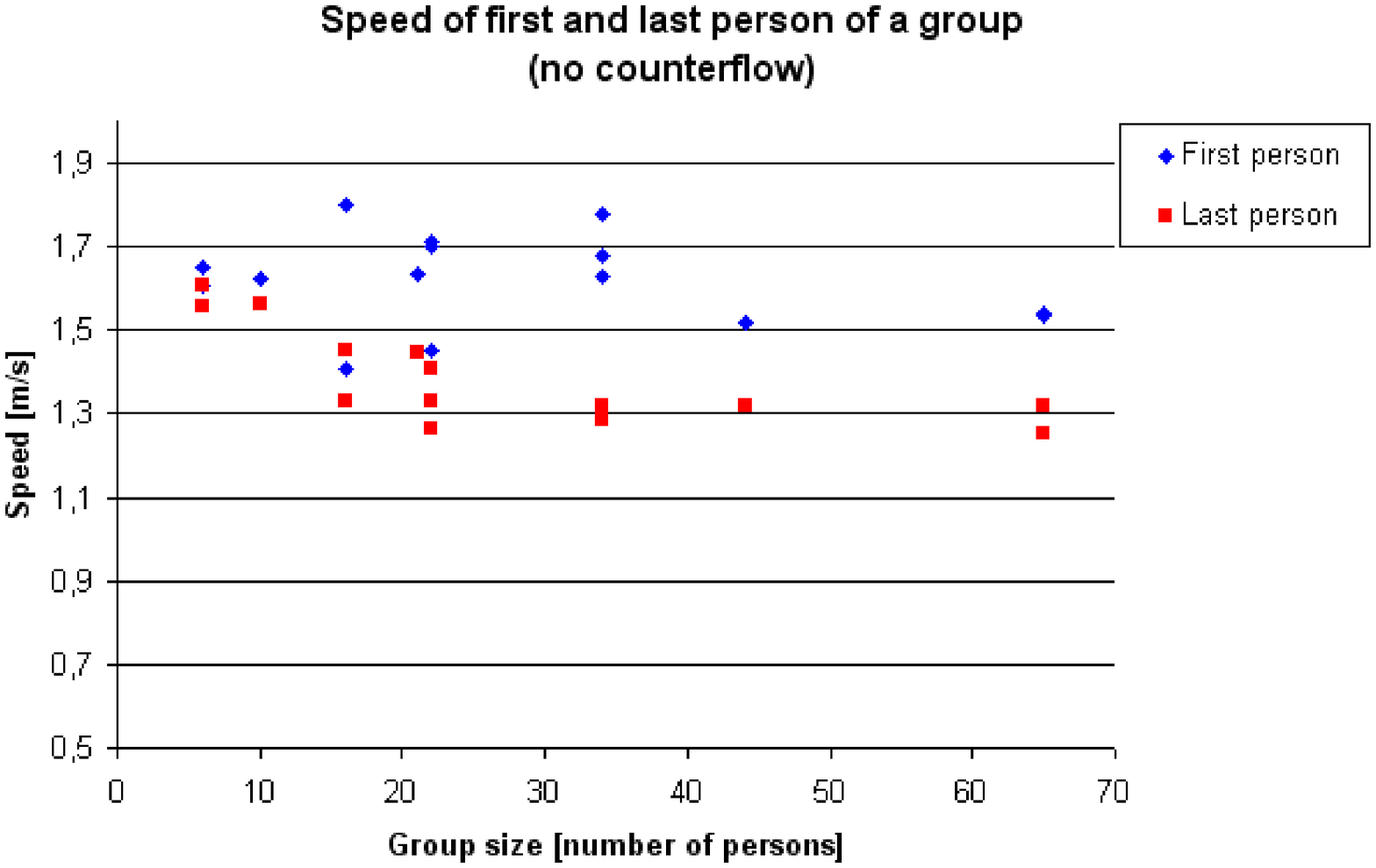}\\ \vspace{12pt}
\includegraphics[width=250pt]{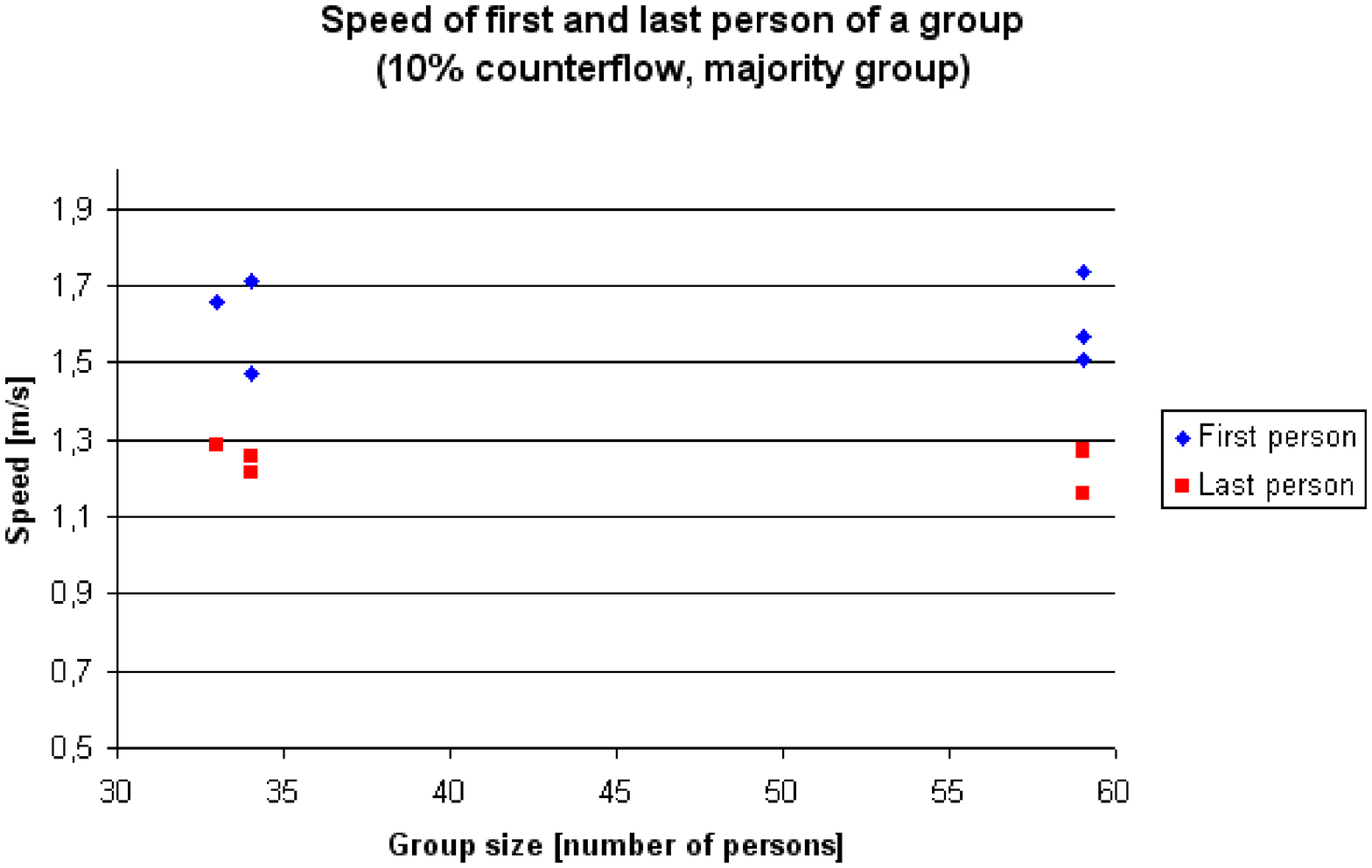}  \\ \vspace{12pt}
\includegraphics[width=250pt]{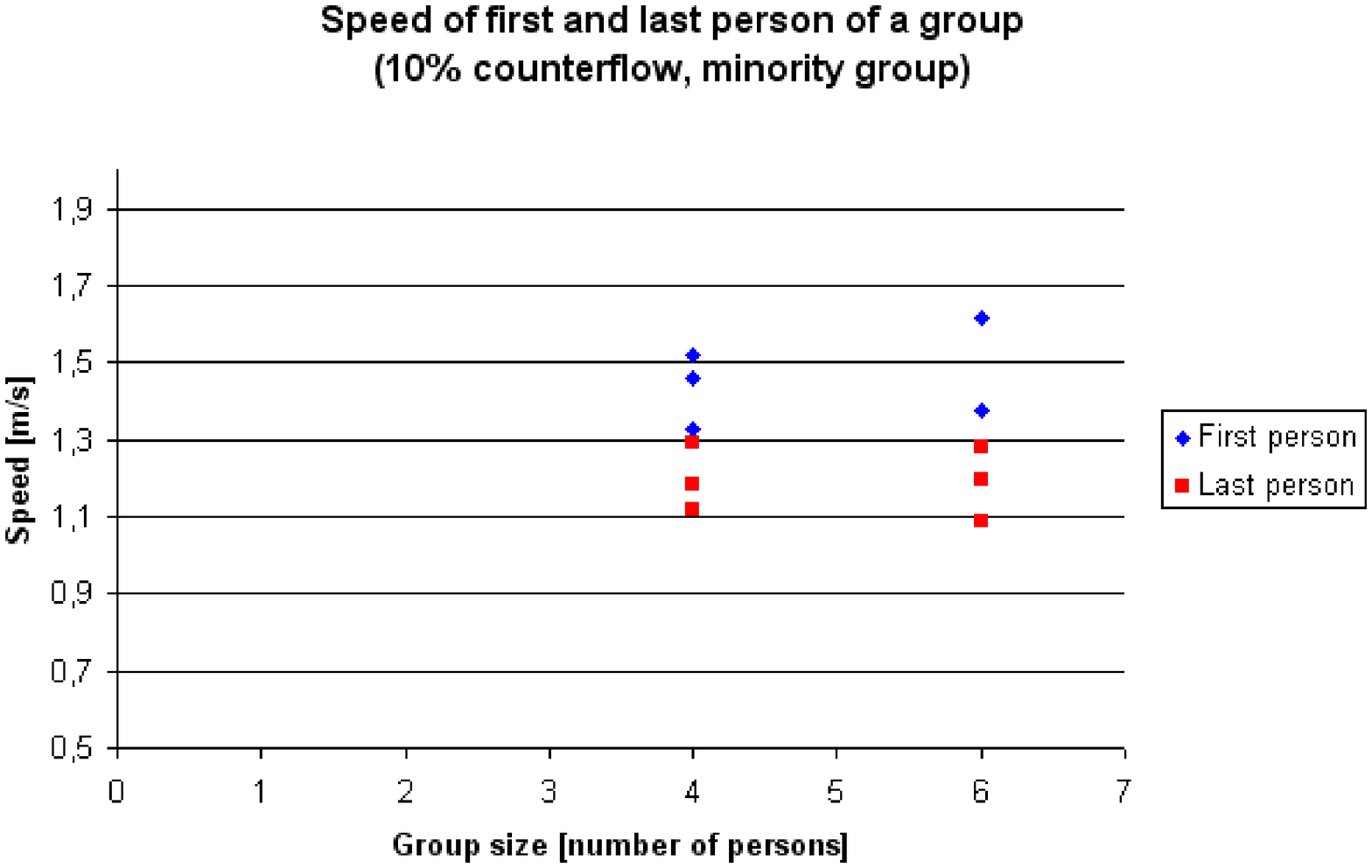} \\
\caption{Walking speeds (part $i$). For no or only small counterflows there is only a small or even no influence of the group size on the walking times.}
\label{fig:speedsA}
\end{center}
\end{figure}

\begin{figure}[htbp]
\begin{center}
\includegraphics[width=250pt]{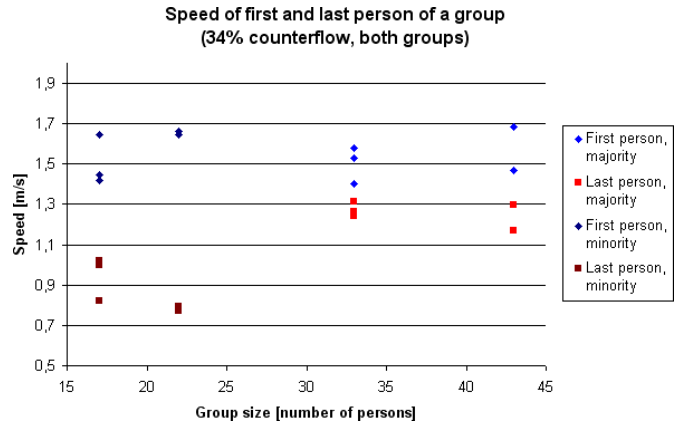}  \\ \vspace{12pt}
\includegraphics[width=250pt]{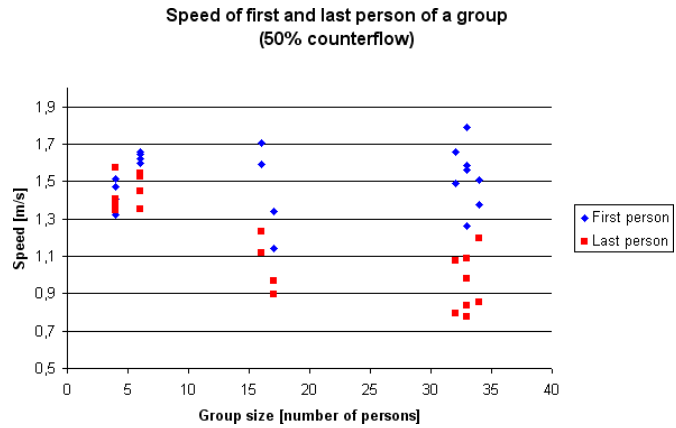} \\
\caption{Walking speeds (part $ii$). The group size has an influence on the walking speed of the last person in the case of $0.5$ counterflow and in terms of the variation of results as well on the speed of the first person.}
\label{fig:speedsB}
\end{center}
\end{figure}

\begin{figure}[htbp]
\begin{center}
\includegraphics[width=250pt]{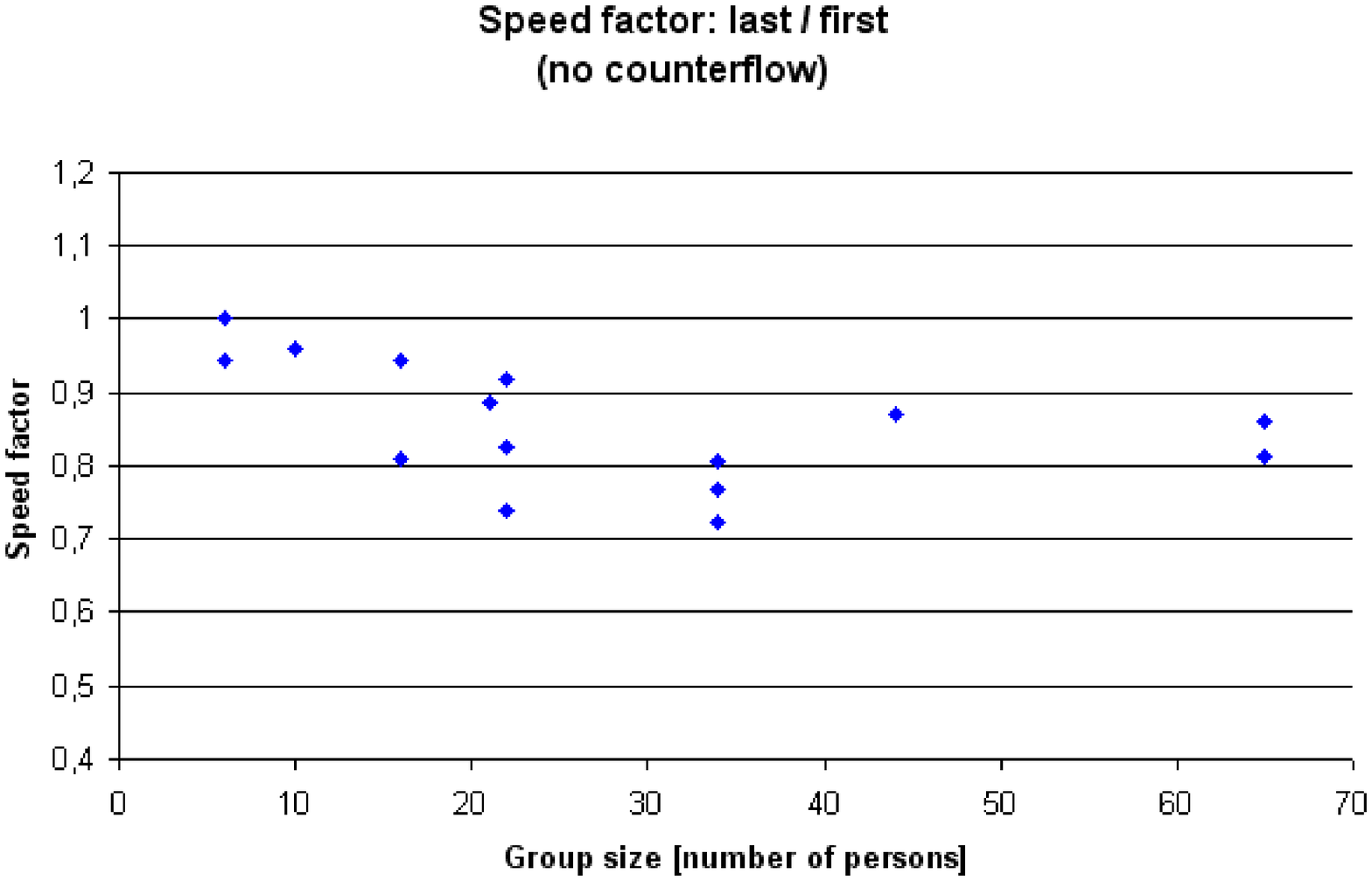} \\ \vspace{12pt}
\includegraphics[width=250pt]{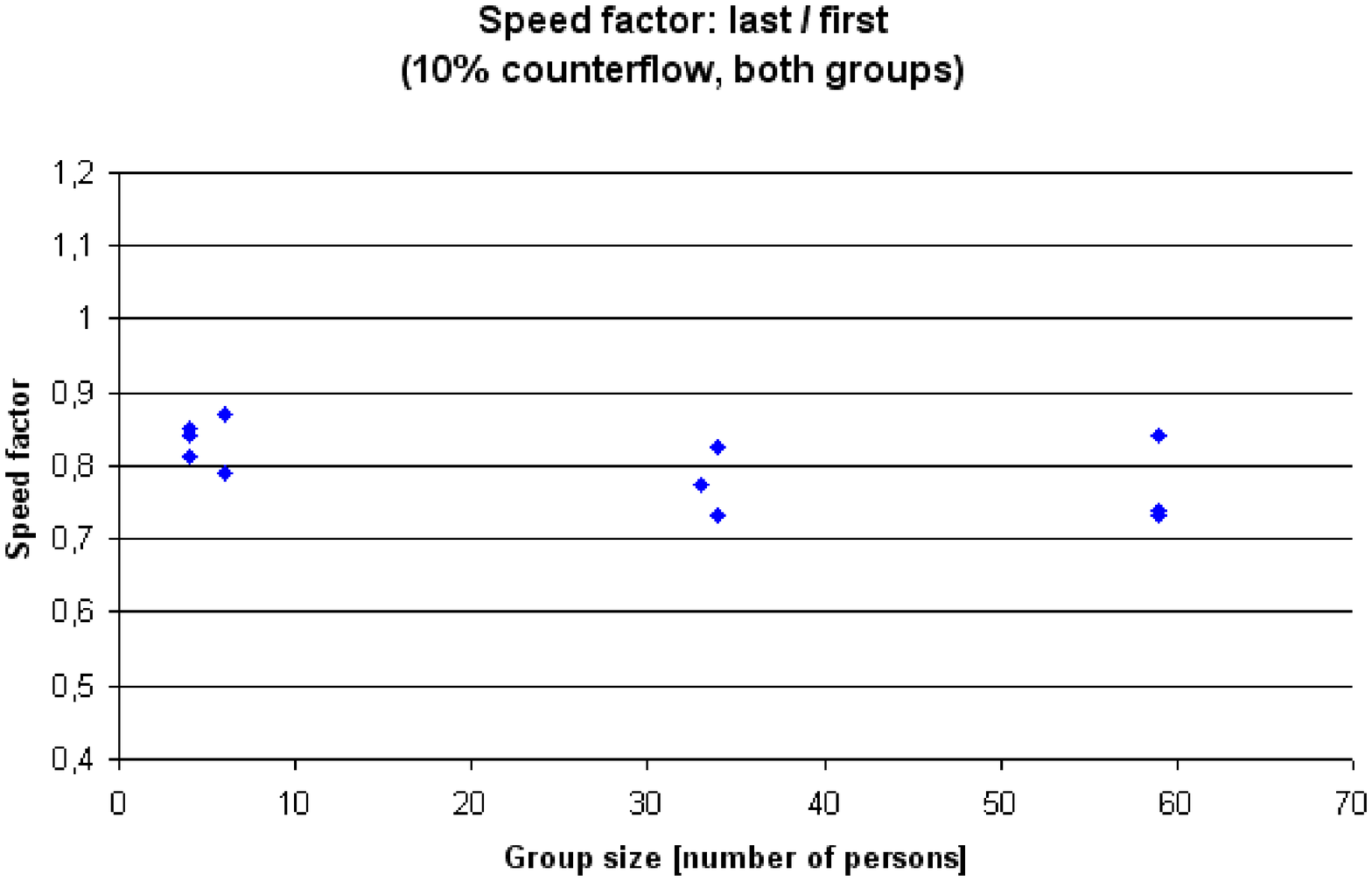} \\
\caption{Walking speed factors (part $i$). This is the ratio of the time that passed from the first person crossing the starting line until passing the finishing line by the same value of the last person. Note that the last and first person may have changed on the way in some cases. For the majority group this ratio is if any then unrecognisably affected bei $0.1$ counterflow compared to no counterflow situations. For the minority group there is an effect denoting that some people (first of the minority group) may do better when walking against a flow than others (last of the minority group).}
\label{fig:speed_factorsA}
\end{center}
\end{figure}

\begin{figure}[htbp]
\begin{center}
\includegraphics[width=250pt]{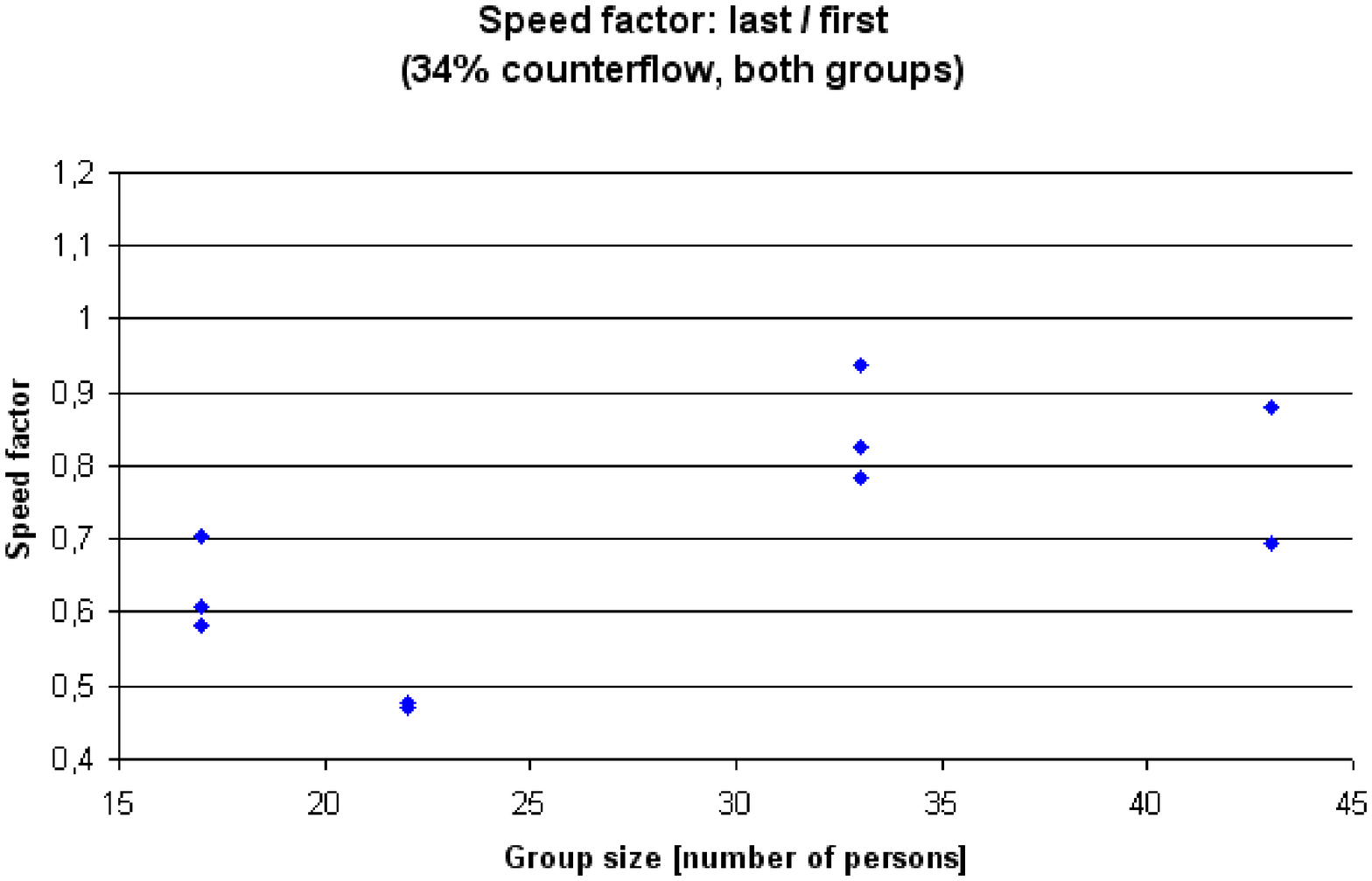} \\ \vspace{12pt}
\includegraphics[width=250pt]{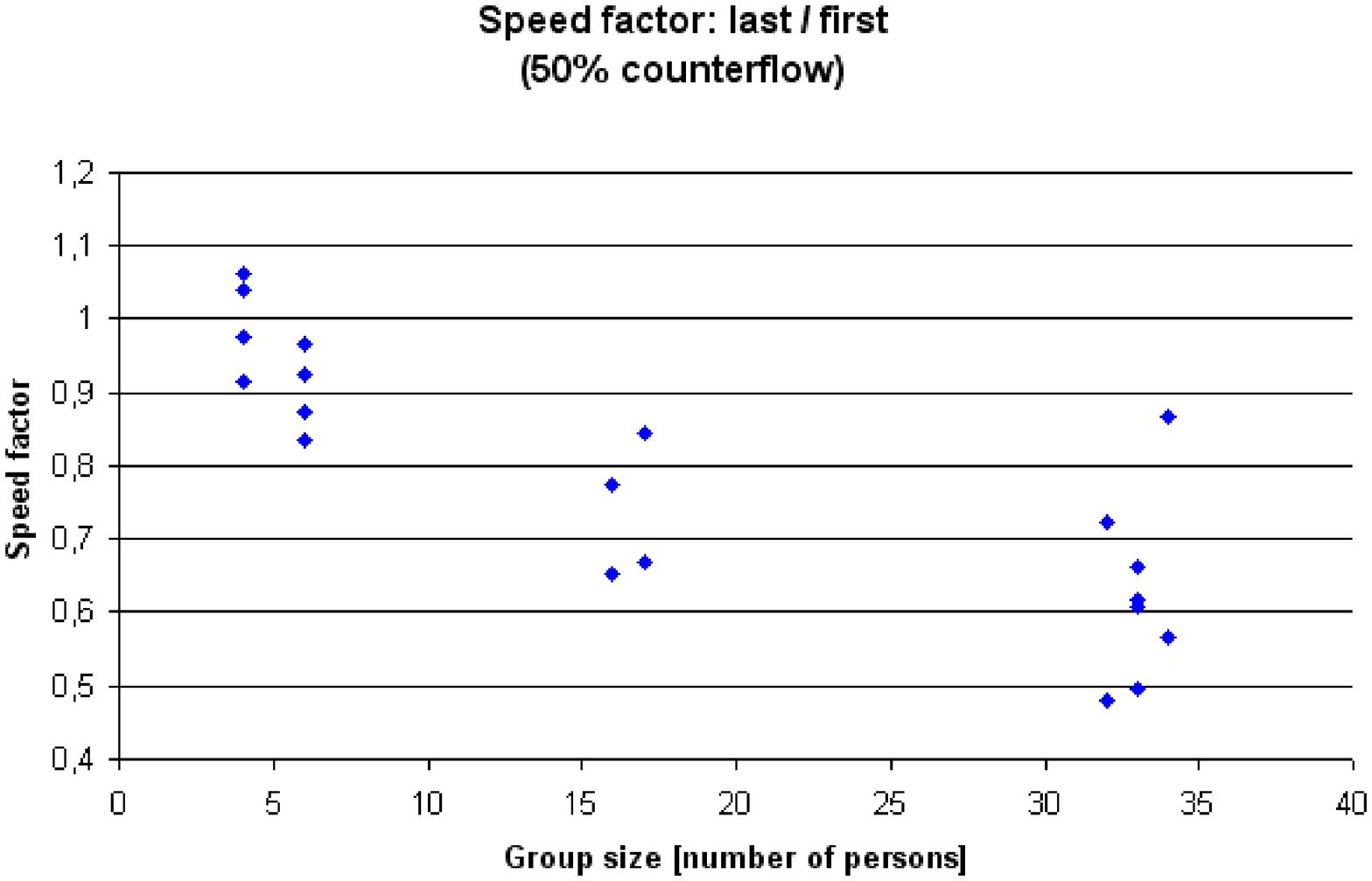} \\
\caption{Walking speed factors (part $ii$). The walking speed factors outline, what has been forshadowed in the walking speeds: the relative difference in the speed of the first and the last person is most distinctive for large groups and $0.5$ counterflow. The figure for $0.34$ counterflow shows, that the minority group in both cases is more affected than the majority group.}
\label{fig:speed_factorsB}
\end{center}
\end{figure}

While the passing time is a time measurement at a certain spot, the walking speeds - to be more precise the speed of the front and the back of a group - relate events at different positions. 
Figures \ref{fig:speedsA} and \ref{fig:speedsB} show the results and figures \ref{fig:speed_factorsA} and \ref{fig:speed_factorsB} the quotients (speed factors) of the walking speeds of the last and first person. 
The speed factors in absence of counterflow never fall below $0.7$, while in counterflow situations they can even be smaller than $0.5$. 
This shows how counterflow situations can loosen walking groups by increasing the distance between the members.
It's an interesting observation that the minority group in a $0.34$ counterflow situation is more affected by this phenomenon than the majority group.
In $0.1$ counterflow situations, however, the minority group seems to be so small (no more than six persons) that they avoid being more loosened than the majority group.
Yet, if one compares the speed factors of the majority group in the $0.1$ counterflow case with those of groups of comparable size in the no counterflow case one finds them to be very similar, while the speed factors of the minority groups are smaller than in cases without counterflow.
This on the contrary could let one conclude that it is the minority group that is affected more than the majority group by the counterflow situation.

\subsection{Fluxes}
\begin{figure}[htbp]
\begin{center}
\includegraphics[width=250pt]{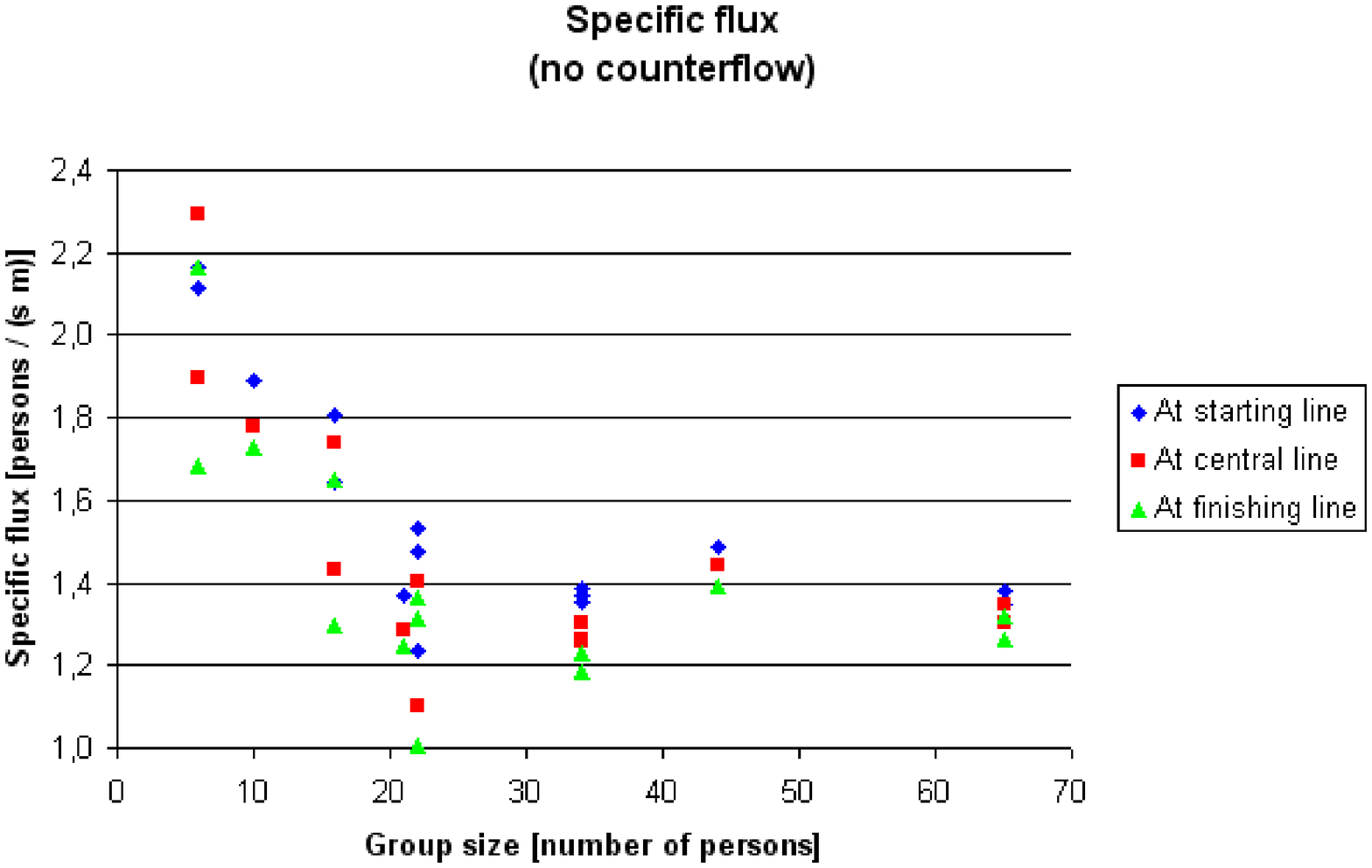}\\ \vspace{12pt}
\includegraphics[width=250pt]{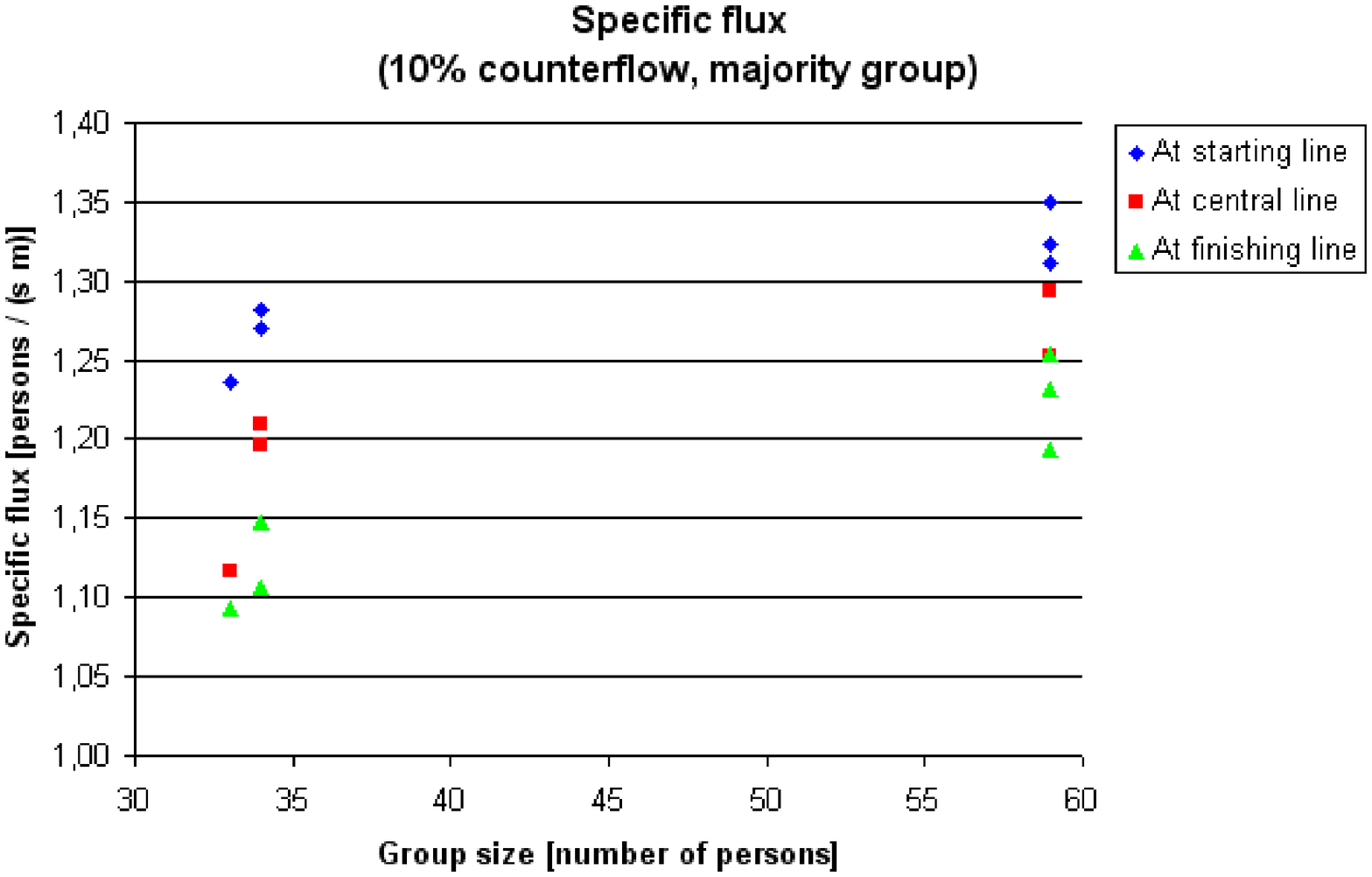}\\ \vspace{12pt}
\includegraphics[width=250pt]{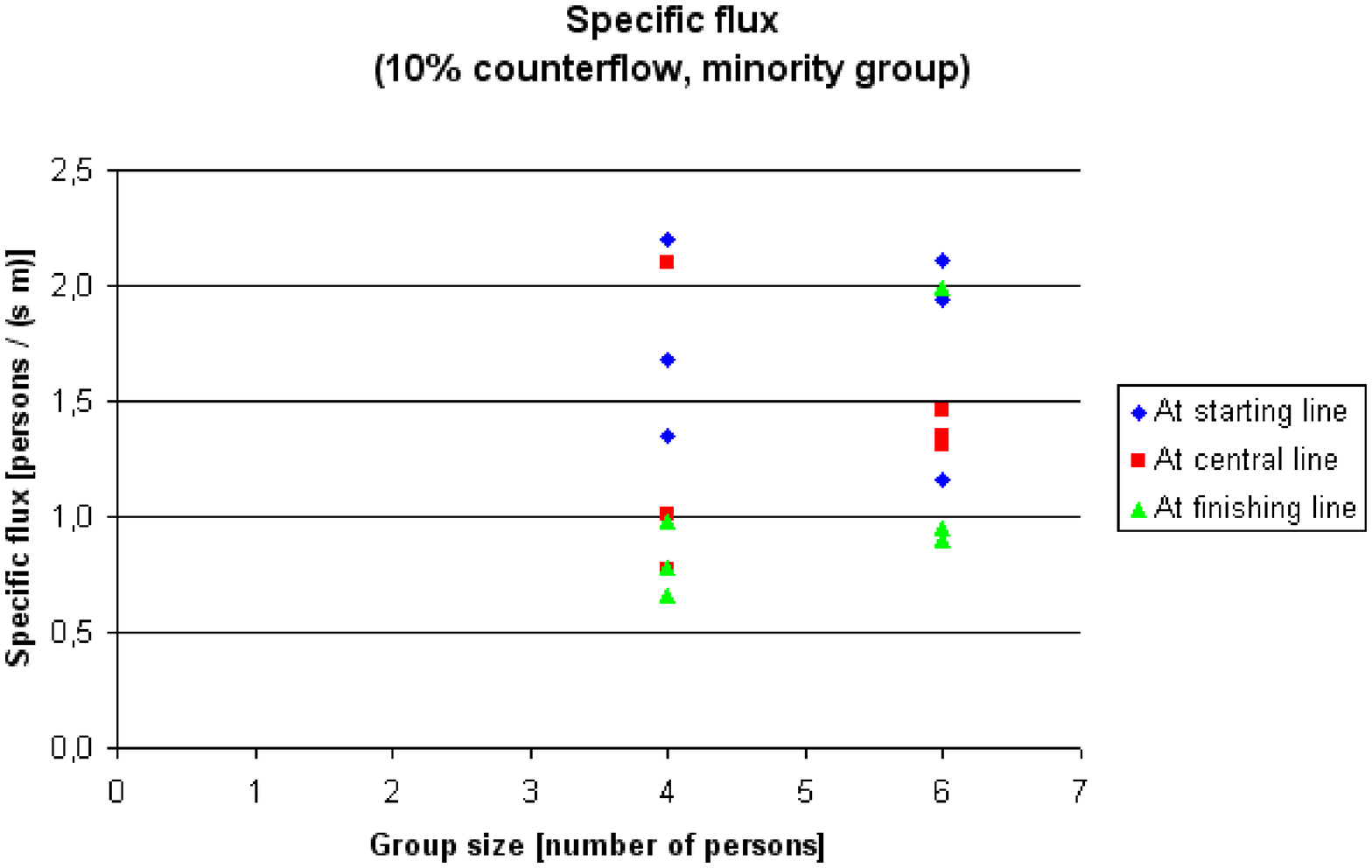} \\ \vspace{12pt}
\caption{Specific fluxes (part $i$): Number of persons / ($1.98$ m $\cdot$ passing time).}
\label{fig:specific_fluxesA}
\end{center}
\end{figure}

\begin{figure}[htbp]
\begin{center}
\includegraphics[width=250pt]{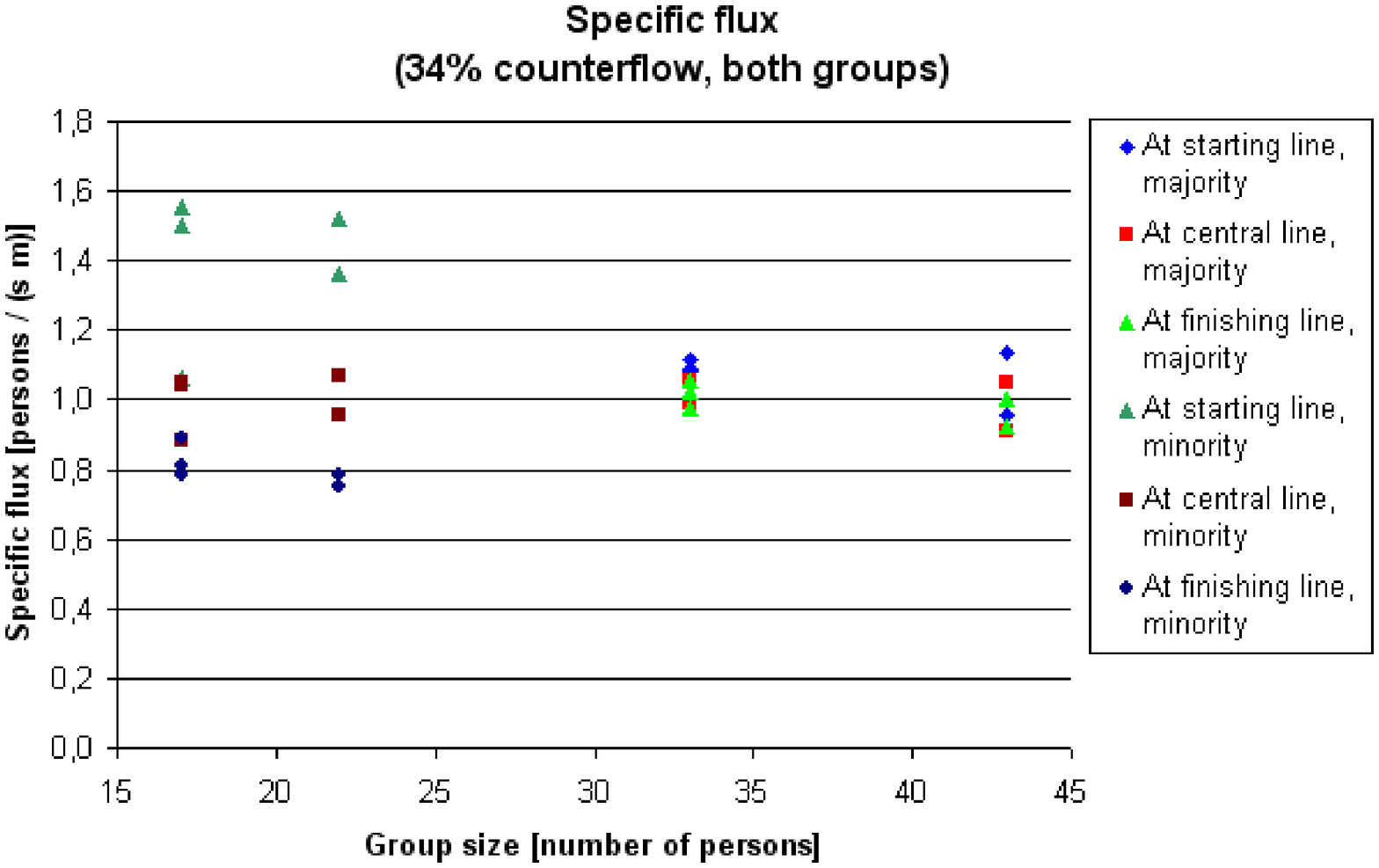}\\ \vspace{12pt}
\includegraphics[width=250pt]{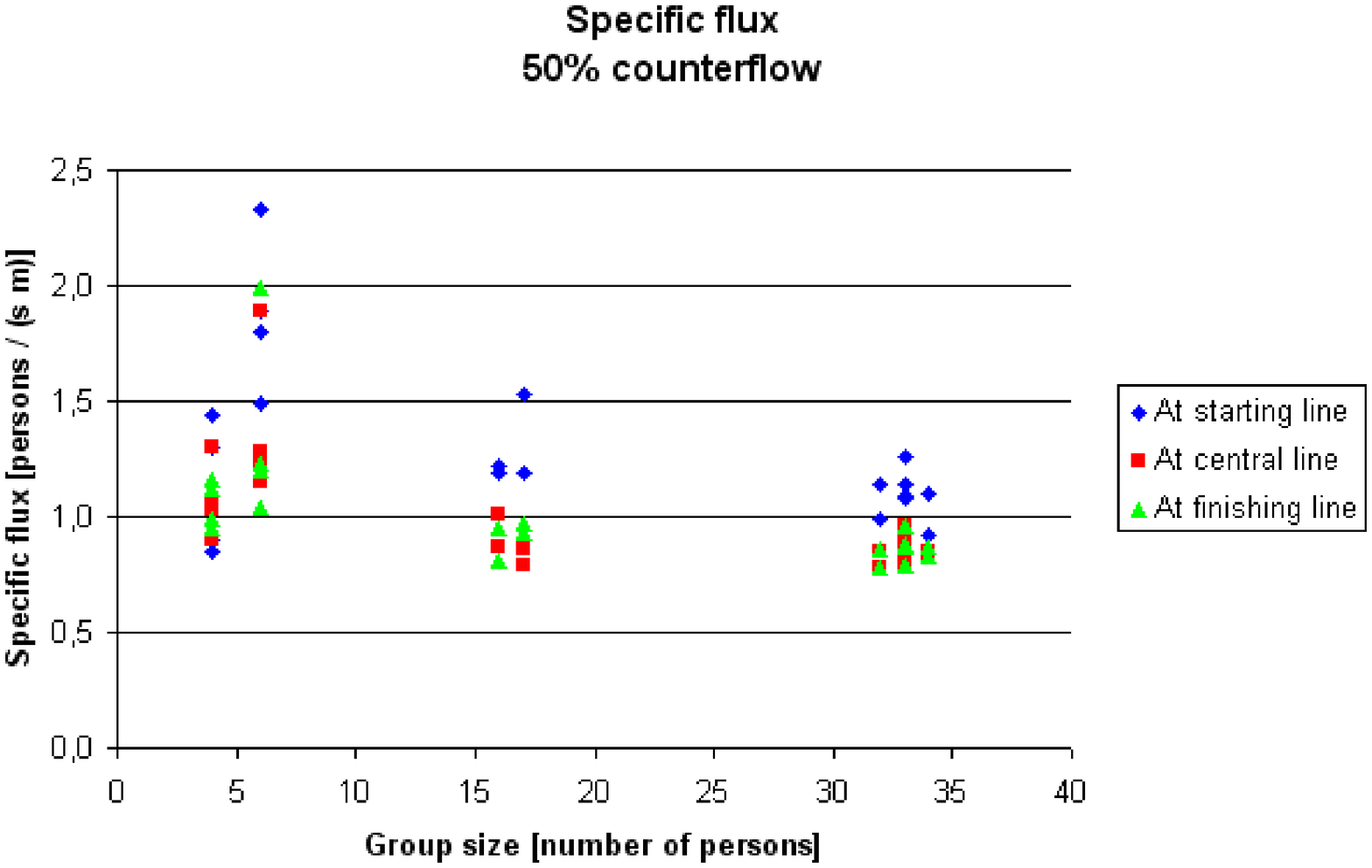} \\
\caption{Specific fluxes (part $ii$): Number of persons / ($1.98$ m $\cdot$ passing time). As with the passing and total times, it also holds for the specific flux, that compared to no counterflow (compare figure \ref{fig:specific_fluxesA}) the performance in the $0.5$ counterflow case is not reduced by a factor of two, but only by approximately a factor of $1.5$ (slightly depending on the measurements taken for comparison). }
\label{fig:specific_fluxesB}
\end{center}
\end{figure}

\begin{figure}[htbp]
\begin{center}
\includegraphics[width=250pt]{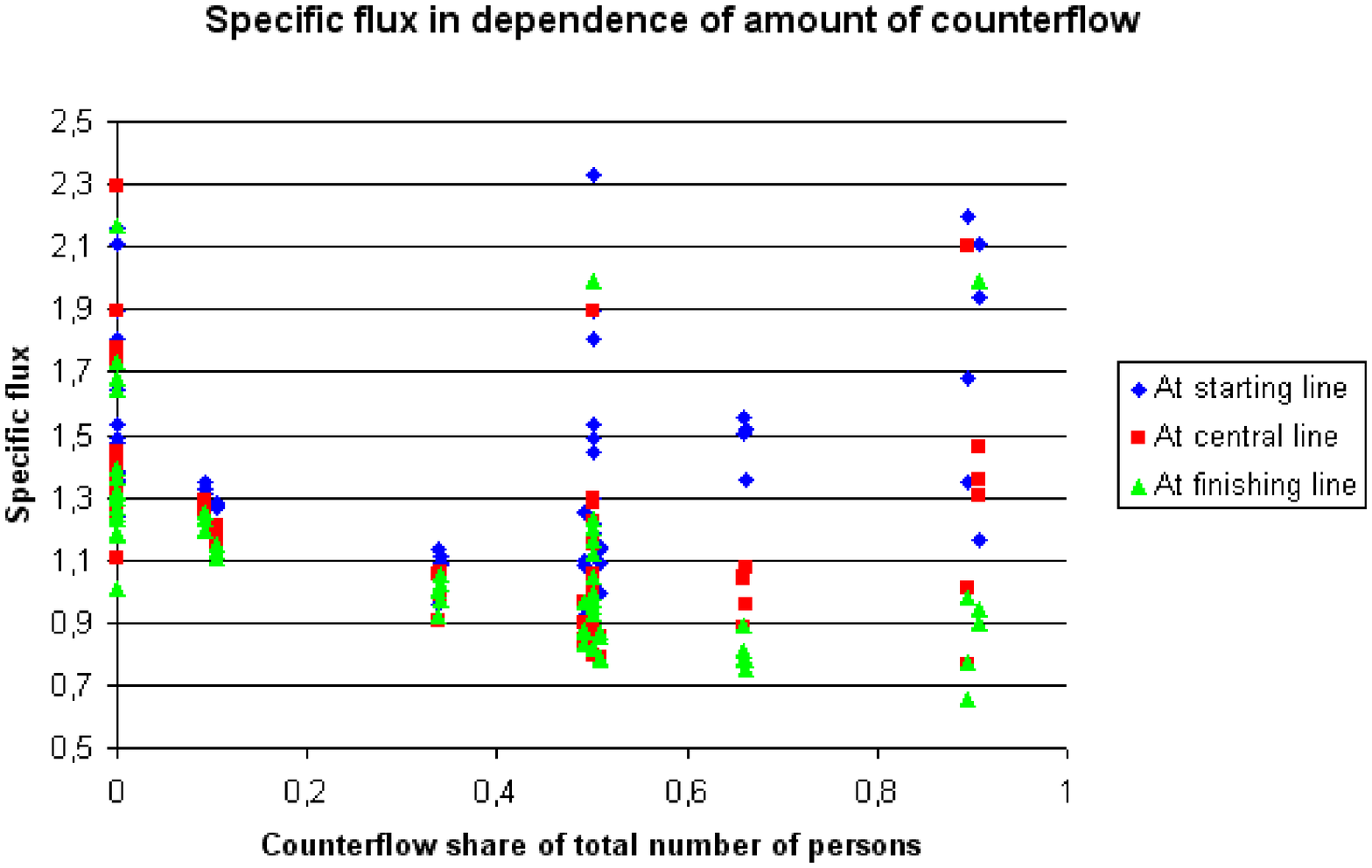}\\ \vspace{12pt}
\includegraphics[width=250pt]{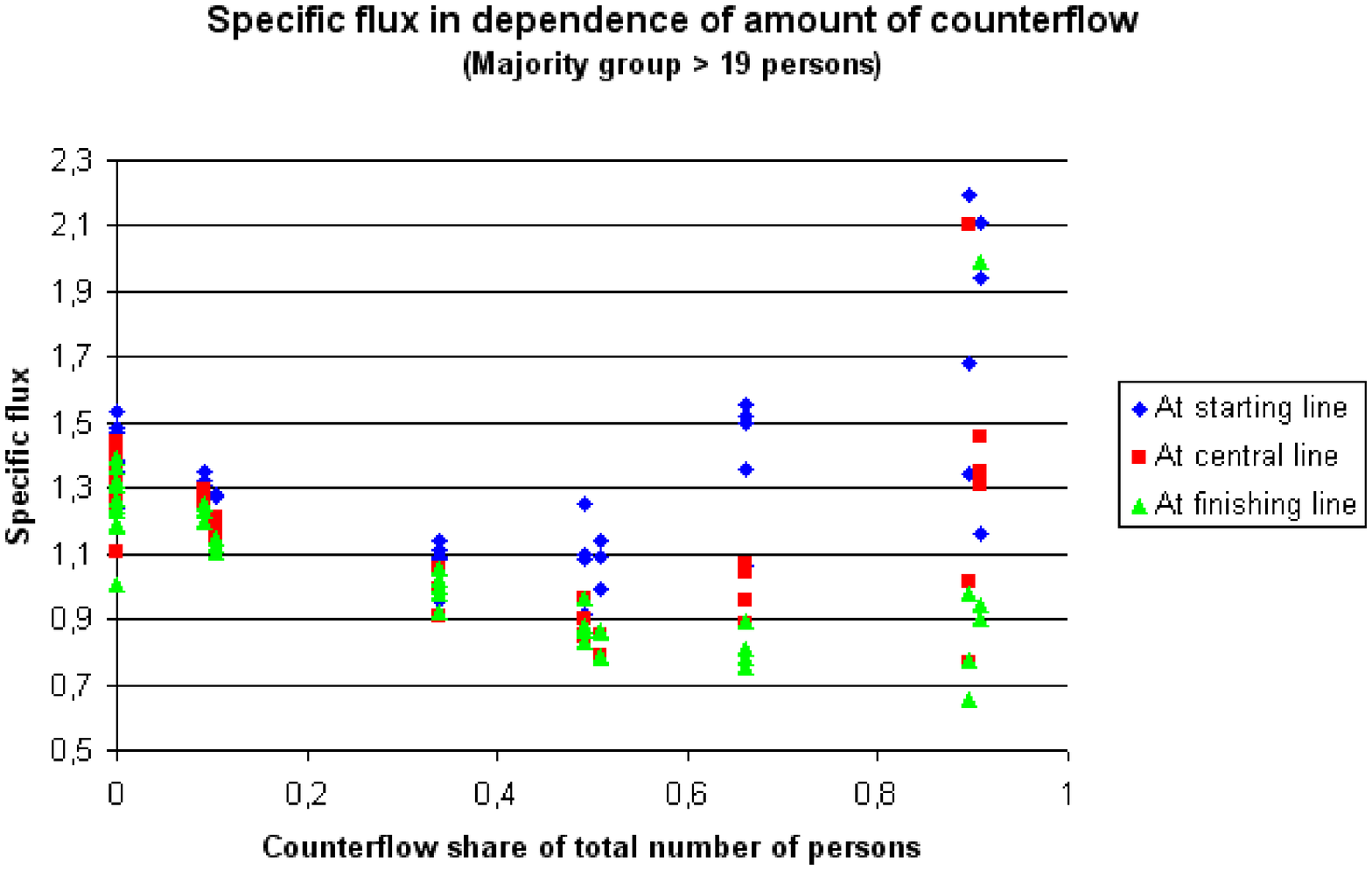}\\ \vspace{12pt}
\includegraphics[width=250pt]{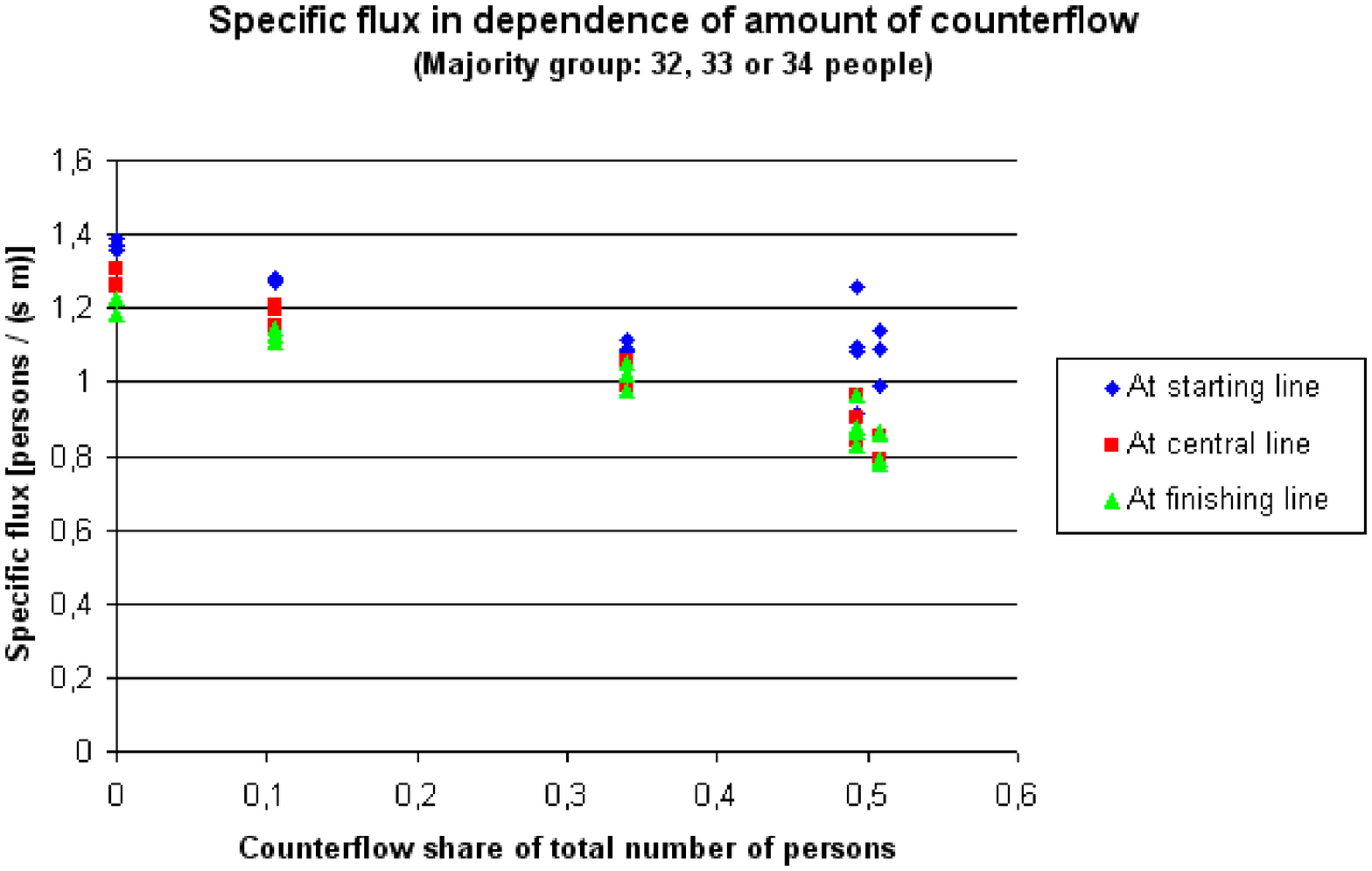}\\
\caption{Comparison of specific fluxes. All results and selections. The third figure shows the results of the largest majority groups that could be formed and significantly less dispersed results than the rest of the data set. At the finishing line the specific flux was found to reduce from approximately $1.2$ (no counterflow) to $0.9$ persons per meter and second ($0.5$ counterflow).}
\label{fig:comp_specific_flux}
\end{center}
\end{figure}

\begin{figure}[htbp]
\begin{center}
\includegraphics[width=250pt]{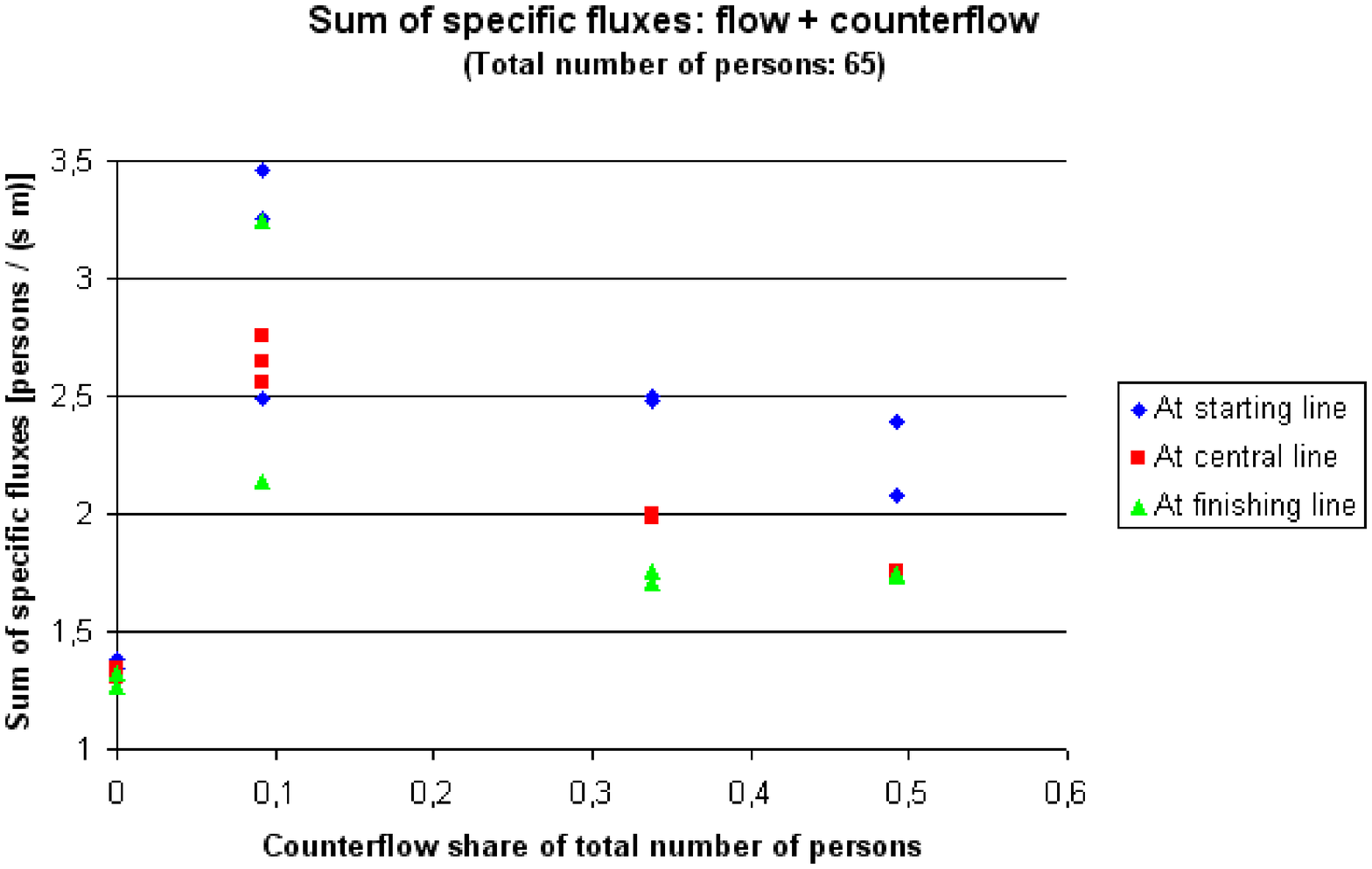}\\
\caption{Comparison of the sum of specific fluxes. The sum of fluxes in counterflow situations is always larger than the flux in the no counterflow situations.}
\label{fig:comp_sum_of_fluxes}
\end{center}
\end{figure}

The specific flux is the number of persons that cross a certain spot divided by the width of the corridor and the (passing) time this process takes. 
A surprising result from figures \ref{fig:specific_fluxesA} and \ref{fig:specific_fluxesB} is that the flux in a $0.5$ counterflow situation is larger than half of the flux in a no counterflow situation with the same total number of persons. (Also compare figures \ref{fig:comp_specific_flux} and \ref{fig:comp_sum_of_fluxes}.)
This is not very surprising at the starting line, i.e. before the two groups meet, as for this spot one effectively has a doubled width compared to no counterflow motion.
But for the central line this is a clear indication that in open-boundary-no-counterflow-situations the density always remains well below the density of maximal flux in the fundamental diagram and that therefore an increased flux is possible for increased density.

If one looks at the specific flux as a function of the fraction of counterflow, one first notices a wide dispersion of the results. 
Only confining the displayed results to those runs where the majority group contained at least $20$ persons shows some clearer trends.
But even then the first trend is that the dispersion of results for the minority group (counterflow fraction $> 0.5$) is large, which might be simply due to the small size of the minority group. (The characteristics of the few individuals who form the group might be crucial.)
Only an experiment with more participants would be able to bring more reliable results for counterflow fractions of $0.9$.
Therefore the impression that the specific flux has a minimum at a counterflow fraction of $0.5$, which could arise from figure \ref{fig:comp_specific_flux}b ,should not be regarded as definite result of this work.
This demands further research, though.\par
A somewhat more evident result is that the flux at the finishing line linearly drops as the fraction of counterflow is increased. (See figure \ref{fig:comp_specific_flux}c.)
With $c_f$ as counterflow fraction the linear regression results in $1.181\cdot persons/(s\cdot m) - 0.374\cdot c_f \cdot persons/(s\cdot m)$ (with $R^2=0.232$) if one only considers runs with a majority group size of at least $20$ people and $1.208\cdot persons/(s\cdot m) - 0.687\cdot c_f \cdot persons/(s\cdot m)$ (with $R^2=0.910$) for runs with a majority group size of $32$, $33$ or $34$ people. 
Here it is interesting to note that the fact that the slope is larger than $-1$ can be interpreted in a way that a counterflow reduces the effective width by a factor smaller than the fraction of counterflow.
This is a similar result as the finding that the flux in a $0.5$ counterflow situation is larger than half of the flux in a no counterflow situation (see above).

\subsection{Lane Formation}
A famous phenomenon in counterflow - but also in some other crowd movement - situations is lane formation \cite{Burstedde01,Helbing95,Hoogendoorn05}.
Pedestrians simply choose to follow closely behind some other person who moves into the same direction.
The lanes that emerge in this way are stable for some time and then disappear, merge or split again.\par
NB: The terminology here is such that a lane can consist of several layers.
Thus if two people can walk side by side without someone in between them who walks in opposite direction, it is still one lane, but two layers.
\subsubsection{Number of Lanes}
\begin{figure}[htbp]
\begin{center}
\includegraphics[width=250pt]{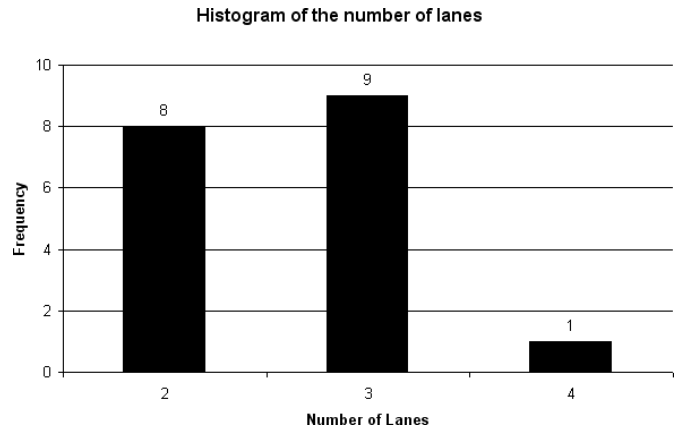}\\
\caption{Histogram of the number of lanes. Only those runs were counted, where the number of lanes could be verified to be the same at CAM C and CAM R (compare figure \ref{fig:floorplan}).}
\label{fig:lanes_hist}
\end{center}
\end{figure}

\begin{figure}[htbp]
\begin{center}
\includegraphics[width=250pt]{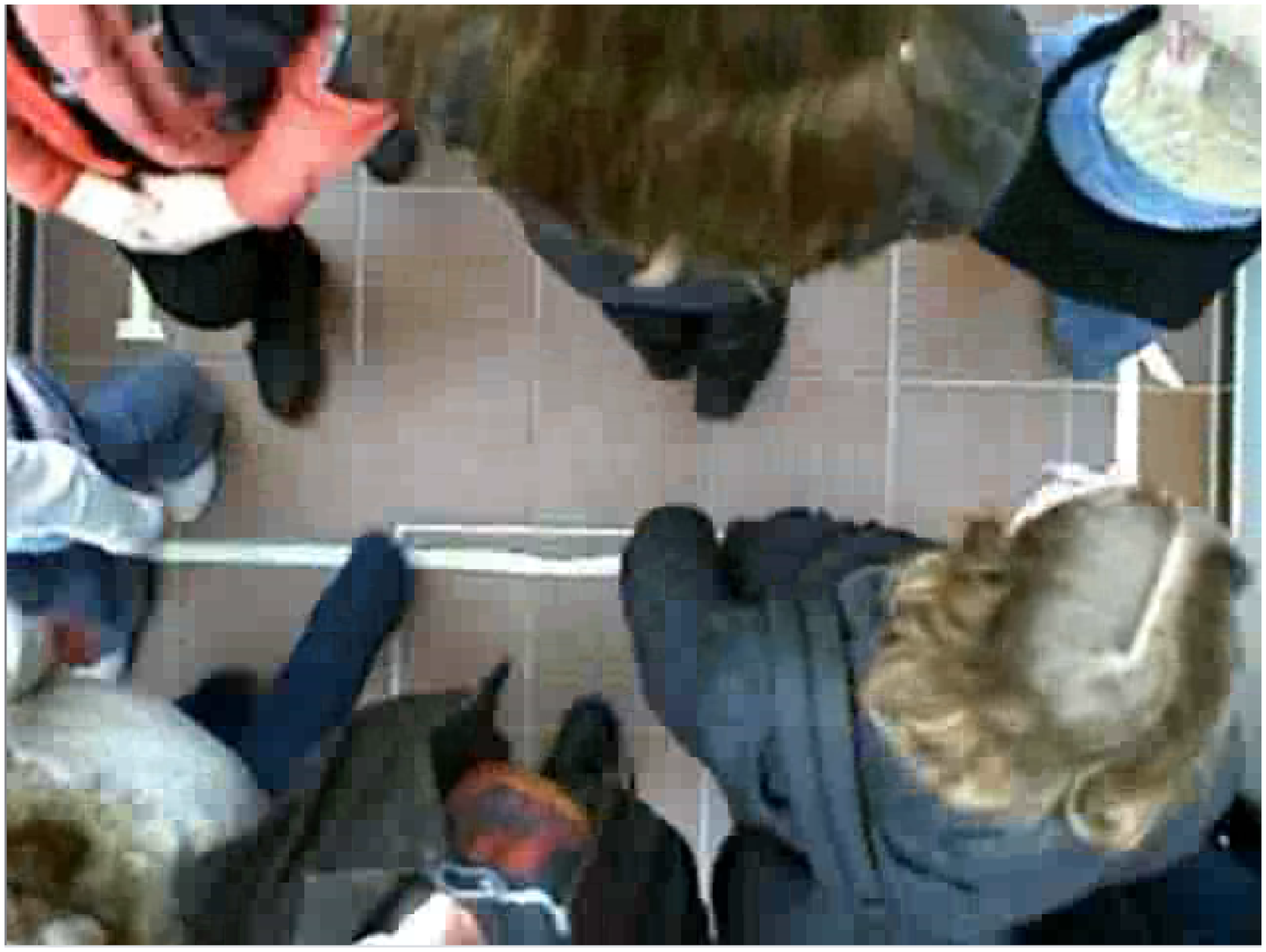}\\ \vspace{12pt}
\includegraphics[width=250pt]{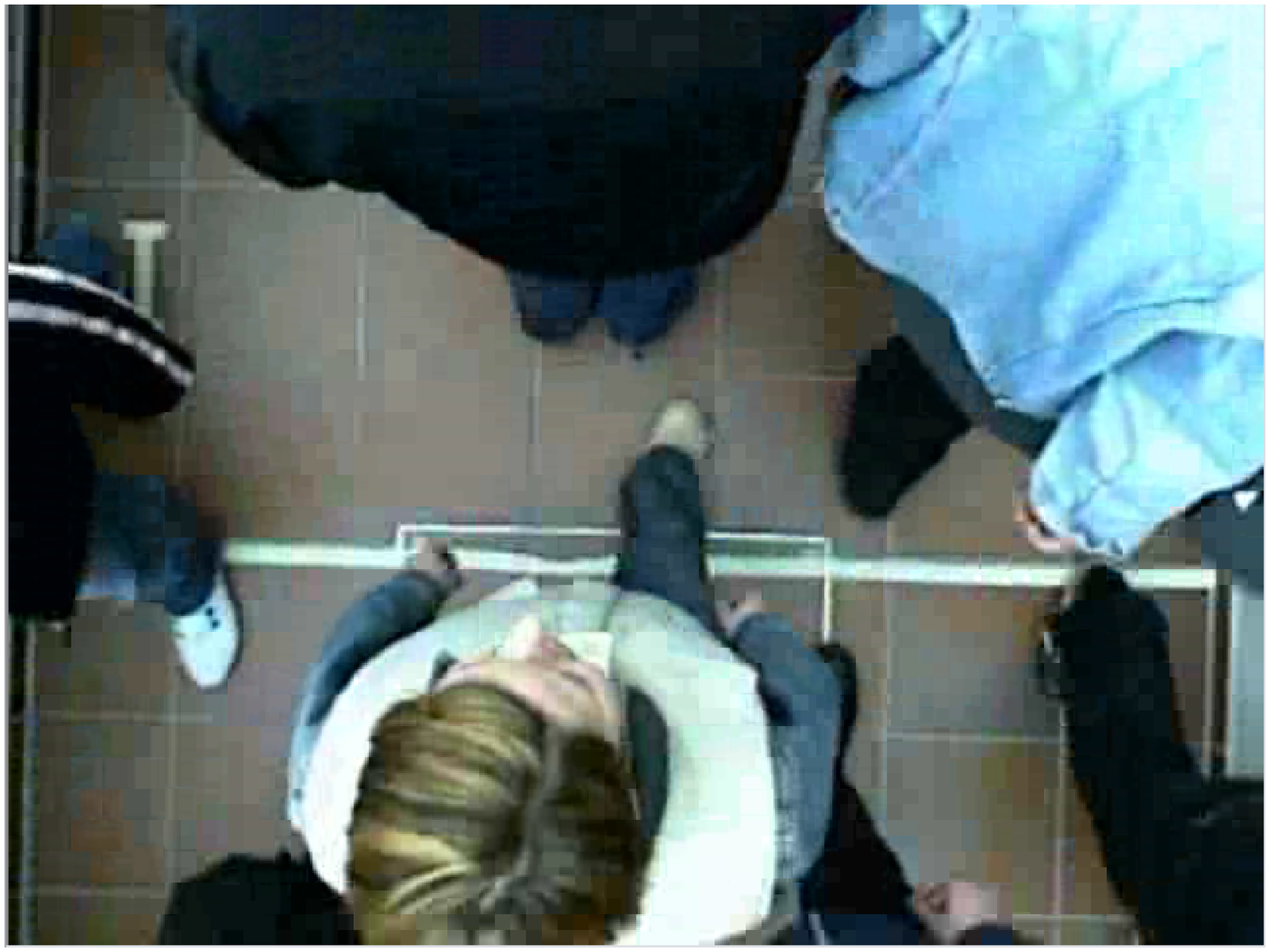}\\ \vspace{12pt}
\includegraphics[width=250pt]{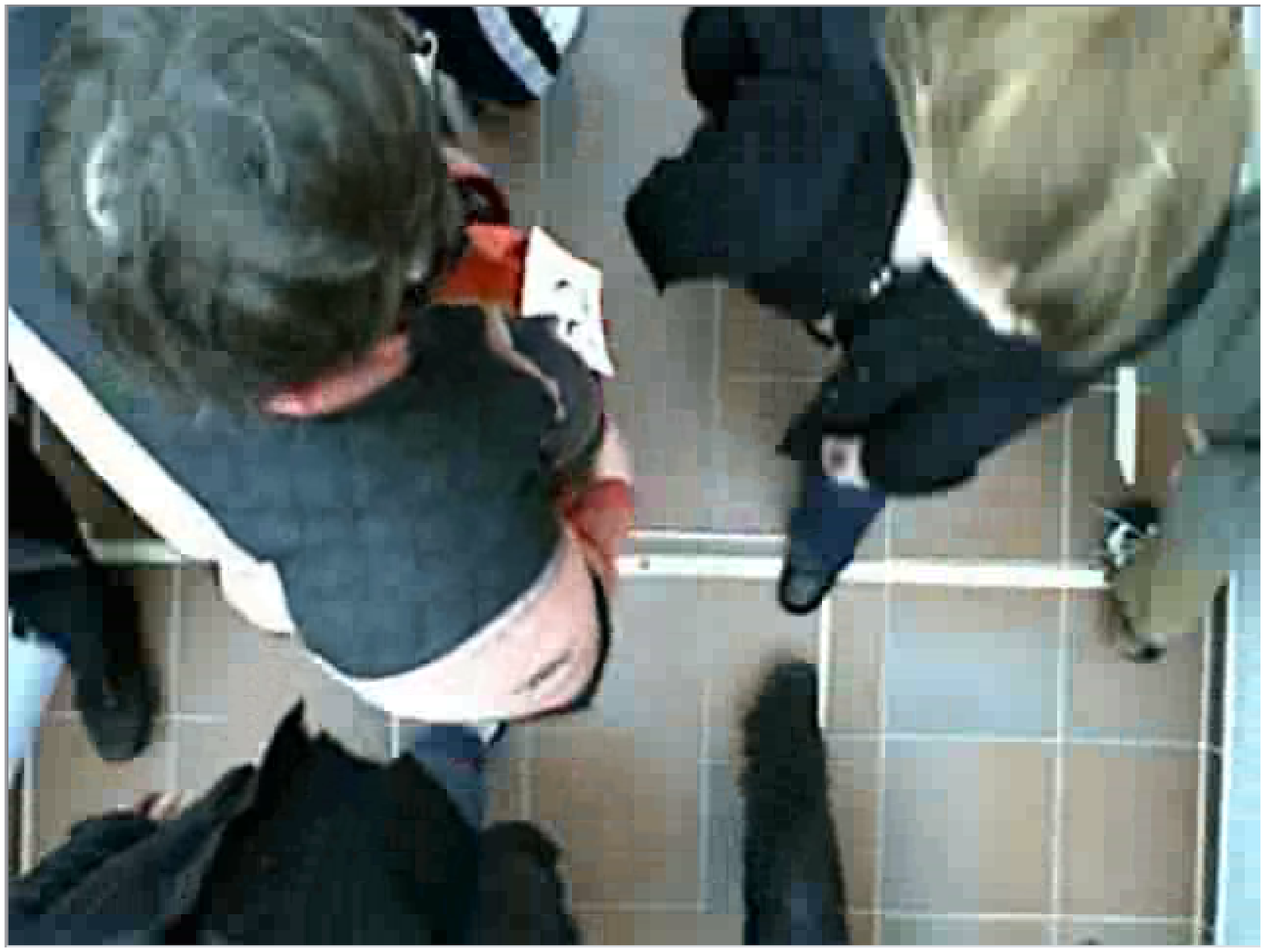}\\
\caption{Two, three and four lanes.}
\label{fig:lanes_pic}
\end{center}
\end{figure}

Figures \ref{fig:lanes_hist} and \ref{fig:lanes_pic} show that in the different runs of this experiment two, three as well as four lanes appeared.
For adults this is probably the maximum of possible cases, as in the case of five lanes there would remain less than $40$ centimeters for each person.

\subsubsection{Left-Right-Asymmetry}
While an odd number of lanes always exhibits a broken symmetry, in the case of an even number of lanes, the symmetry could be preserved, as long as right-hand traffic and left-hand traffic appear in equal shares. 
In the nine cases of an even number of lanes, however, left-hand traffic did not appear even once. 
The participants always ``chose" right-hand traffic.

\section{Summary, Conclusions and Outlook}
In this work the results of a counterflow in a corridor experiment are presented. 
The results yield comparatively high speeds and fluxes, which probably is mainly due to the participants mostly being in their twenties. 
Compared to a situation without counterflow the performance - in terms of passing or total times, speed, and flux - of a group of walkers is never reduced as much as one would expect from the amount of counterflow. 
For example if there are two equally sized groups the passing time at a certain spot does not double, nor does the flux drop to $50$\%.
However, the passing time increases and the flux decreases, the participants seemed to be able to compensate the existence of a counterflow to a certain point by accepting higher densities and using space more efficiently.
This phenomenon can be summed up by saying that the sum of fluxes in a counterflow situation in this experiment was always found to be larger than the flux in any of the no counterflow situations.
Implementing this high efficiency of real pedestrians to realistic simulation models of pedestrian dynamics will pose a challenge, as it seems that a lot of details and a significant amount of intellectual power have to be considered.
\par
A maximal asymmetry between right- and left-hand traffic was found and in one run as much as four lanes.\par
Some of the results - mainly those with small fractions of counterflow larger zero - demand confirmation in experiments with more participants.
This would in addition grant the possibility to evaluate the time evolution of the observables (flux, speed, stability of lanes).
It would furthermore be very interesting to repeat the experiment 
1) with participants mostly older than $60$ years; 
2) in corridors with different widths, e.g., $1.5$, $2.5$, $3.0$ or $5.0$ meters; 
3) in a country with left-hand traffic, to check the correlation between car traffic rule's and pedestrian behaviour; and 
4) in a ring shaped corridor for cyclic boundary conditions and therefore constant global density.\par

\subsection*{Acknowledgments}
This work has been financed by the ``Bundesministerium f\"ur Bildung und Forschung" (BMBF) within the project PeSOS . We thank Mr. Baues, headmaster the Sportschule Wedau, for allowing us to use the building for the experiment, Christian Ehling, Lars Habel, Hubert Kl\"upfel, and Mareike Quessel for being very helpful in the conduction of the experiment, Hubert Kl\"upfel and Mareike Quessel especially for documenting the experiment photographically respectively for creating the floor plan (figure \ref{fig:floorplan}) and of course all of the participants.

\bibliographystyle{unsrt}
\bibliography{counterflow_corridor}
\nocite{PED01}

\end{document}